\documentclass[aps,prx,twocolumn,superscriptaddress,longbibliography,floatfix]{revtex4-2}
\usepackage{amsmath,amssymb,amsfonts}
\usepackage{xcolor,graphicx}

\usepackage[
  bookmarks=true,
  bookmarksnumbered=true,
  bookmarksopen=true,
  pdfpagemode=UseOutlines,
  linkcolor=cyan,
  urlcolor=blue,
  colorlinks,
  citecolor=cyan
]{hyperref}

\makeatletter
\def\NAT@sort{\z@}
\makeatother

\begin{document}

\title{Connectivity-induced surface-loss penalty
in superconducting qubit-coupler lattices}

\author{Xu-Yang Gu}
\affiliation{Beijing National Laboratory for Condensed Matter Physics, Institute of Physics, Chinese Academy of Sciences, Beijing 100190, China}
\affiliation{School of Physical Sciences, University of Chinese Academy of Sciences, Beijing 100049, China}

\author{Gui-Han Liang}
\affiliation{Beijing National Laboratory for Condensed Matter Physics, Institute of Physics, Chinese Academy of Sciences, Beijing 100190, China}

\author{Ming-Chuan Wang}
\affiliation{Beijing National Laboratory for Condensed Matter Physics, Institute of Physics, Chinese Academy of Sciences, Beijing 100190, China}
\affiliation{School of Physical Sciences, University of Chinese Academy of Sciences, Beijing 100049, China}

\author{Yongxi Xiao}
\affiliation{Beijing National Laboratory for Condensed Matter Physics, Institute of Physics, Chinese Academy of Sciences, Beijing 100190, China}
\affiliation{School of Physical Sciences, University of Chinese Academy of Sciences, Beijing 100049, China}

\author{Cheng-Lin Deng}
\affiliation{Beijing Key Laboratory of Fault-Tolerant Quantum Computing, Beijing Academy of Quantum Information Sciences, Beijing 100193, China}

\author{Zheng-He Liu}
\affiliation{Beijing National Laboratory for Condensed Matter Physics, Institute of Physics, Chinese Academy of Sciences, Beijing 100190, China}
\affiliation{School of Physical Sciences, University of Chinese Academy of Sciences, Beijing 100049, China}

\author{Tian-Ming Li}
\affiliation{Beijing National Laboratory for Condensed Matter Physics, Institute of Physics, Chinese Academy of Sciences, Beijing 100190, China}
\affiliation{School of Physical Sciences, University of Chinese Academy of Sciences, Beijing 100049, China}

\author{Kai Xu}
\email{kaixu@iphy.ac.cn}
\affiliation{Beijing National Laboratory for Condensed Matter Physics, Institute of Physics, Chinese Academy of Sciences, Beijing 100190, China}
\affiliation{School of Physical Sciences, University of Chinese Academy of Sciences, Beijing 100049, China}
\affiliation{Beijing Key Laboratory of Fault-Tolerant Quantum Computing, Beijing Academy of Quantum Information Sciences, Beijing 100193, China}
\affiliation{Hefei National Laboratory, Hefei 230088, China}
\affiliation{Beijing Key Labratory of Advanced Quantum Technology, Beijing 100190, China}

\author{Zhongcheng Xiang}
\email{zcxiang@iphy.ac.cn}
\affiliation{Beijing National Laboratory for Condensed Matter Physics, Institute of Physics, Chinese Academy of Sciences, Beijing 100190, China}
\affiliation{School of Physical Sciences, University of Chinese Academy of Sciences, Beijing 100049, China}
\affiliation{Hefei National Laboratory, Hefei 230088, China}
\affiliation{Beijing Key Labratory of Advanced Quantum Technology, Beijing 100190, China}

\author{Heng Fan}
\email{hfan@iphy.ac.cn}
\affiliation{Beijing National Laboratory for Condensed Matter Physics, Institute of Physics, Chinese Academy of Sciences, Beijing 100190, China}
\affiliation{School of Physical Sciences, University of Chinese Academy of Sciences, Beijing 100049, China}
\affiliation{Beijing Key Laboratory of Fault-Tolerant Quantum Computing, Beijing Academy of Quantum Information Sciences, Beijing 100193, China}
\affiliation{Hefei National Laboratory, Hefei 230088, China}
\affiliation{Beijing Key Labratory of Advanced Quantum Technology, Beijing 100190, China}


\begin{abstract}
Recent advances in design and fabrication
have increased the energy-relaxation times of 
isolated superconducting transmon qubits to 
the hundreds-of-microseconds regime, 
with reported values exceeding 500~$\mu$s.  
However, the same progress has not automatically
translated to multiqubit processors, 
where qubits are embedded in connected qubit-coupler lattices 
and often exhibit much shorter lifetimes than isolated qubits.    
To identify possible sources of this discrepancy,
here we use finite-element simulation to investigate 
how surface participation ratios and
the resulting surface dielectric loss change 
when a qubit is embedded in a flip-chip qubit-coupler lattice.  
Controlled comparisons show that 
higher connectivity can indeed lead to larger surface loss: 
in the simulated lattice, connecting a qubit to two and four couplers
increases the surface loss by factors of 1.3 and 1.8, respectively.  
We attribute this change to the combined effects of 
added edge fields from coupling claws, 
field redistribution over the larger connected metal network,
and hybridization with coupler modes.  
We further examine how this connectivity-induced surface-loss penalty 
depends on the
geometric design parameters 
of both the qubit electrodes and the coupling claws, 
and derive guidelines for designing low-loss multiqubit processors.
\end{abstract}
\maketitle

\section{Introduction}
\label{sec:intro}

Superconducting qubits are among the leading platforms for
realizing fault-tolerant quantum computation, 
as demonstrated in recent works
~\cite{zhaoRealizationErrorCorrectingSurface2022,
googlequantumaiSuppressingQuantumErrors2023,
googleQuantumAIQuantumErrorCorrection2025,
lacroixScalingLogicColour2025,wangDemonstrationLowoverheadQuantum2026a}.
To advance from these lab-scale demonstrations to a practical
fault-tolerant quantum computer, several engineering challenges must
be addressed.  One central requirement is to suppress decoherence in
superconducting qubits while increasing the number of qubits integrated
on a chip.

Substantial progress has been made by identifying and reducing
relaxation channels in transmon qubits, including dielectric
loss~\cite{wennerSurfaceLossSimulations2011a,wangSurfaceParticipationDielectric2015,dialBulkSurfaceLoss2016a,gambettaInvestigatingSurfaceLoss2017,crowleyDisentanglingLossesTantalum2023a,ganjamSurpassingMillisecondCoherence2024},
Purcell decay~\cite{houckControllingSpontaneousEmission2008,sheldonCharacterizationHiddenModes2017},
and quasiparticle noise~\cite{risteMillisecondChargeparityFluctuations2013,serniakHotNonequilibriumQuasiparticles2018a,diamondDistinguishingParitySwitchingMechanisms2022,mcewenResistingHighEnergyImpact2024a,glazmanBogoliubovQuasiparticlesSuperconducting2021,catelaniUsingMaterialsQuasiparticle2022}.
Through optimized designs and fabrication processes that reduce
dielectric loss~\cite{martinisUCSBFinalReport2014,dunsworthCharacterizationReductionCapacitive2017a,murrayMaterialMattersSuperconducting2021,placeNewMaterialPlatform2021,martinisSurfaceLossCalculations2022,crowleyDisentanglingLossesTantalum2023a,blandMillisecondLifetimesCoherence2025b,biznarovaMitigationInterfacialDielectric2024,mahuliImprovingLifetimeAluminumBased2025},
multistage shielding to suppress quasiparticle
noise~\cite{barendsMinimizingQuasiparticleGeneration2011,corcolesProtectingSuperconductingQubits2011,gordonEnvironmentalRadiationImpact2022},
and optimization of the on-chip~\cite{reedFastResetSuppressing2010a,jeffreyFastAccurateState2014,chenFabricationCharacterizationAluminum2014}
and off-chip~\cite{wennerWirebondCrosstalkCavity2011,huangMicrowavePackageDesign2021}
electromagnetic environments, the relaxation times of isolated
transmon qubits have increased from the few- to few-tens-of-microseconds
scale to the hundreds-of-microseconds regime
\cite{placeNewMaterialPlatform2021,wangPracticalQuantumComputers2022,
gordonEnvironmentalRadiationImpact2022,
balSystematicImprovementsTransmon2024,
biznarovaMitigationInterfacialDielectric2024,
tuokkolaMethodsAchieveNearmillisecond2025,
blandMillisecondLifetimesCoherence2025b}.

\begin{figure*}[t]
    \centering
    \includegraphics[width=0.92\textwidth]{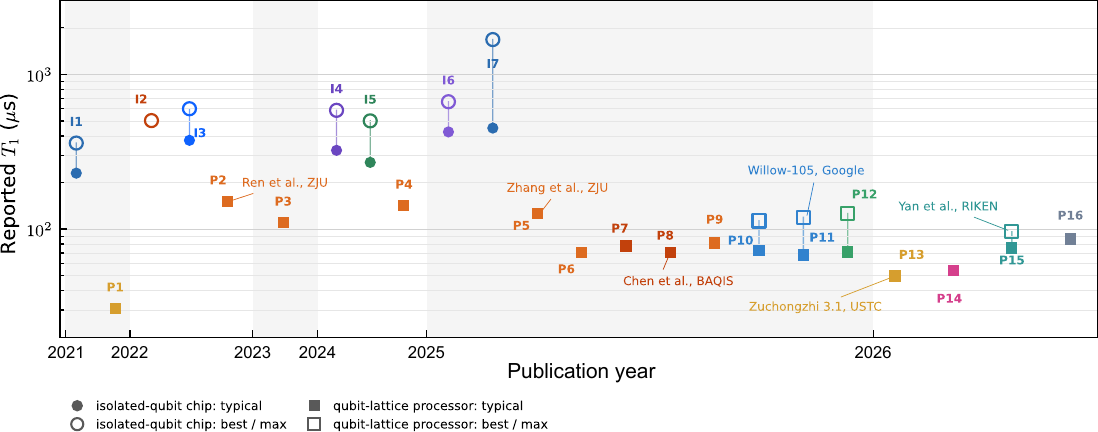}
    \caption{
    Comparison between the reported energy-relaxation times of isolated-qubit chips 
    and qubit-lattice processors.
    Filled symbols denote reported typical values such as mean,
    median, or average $T_1$, and open symbols denote reported best or maximum
    values when available.  
    Marker shape separates isolated-qubit chips from
    qubit-lattice processors.
    The plotted isolated-qubit chips are taken from
    Refs.~\cite{placeNewMaterialPlatform2021,wangPracticalQuantumComputers2022,
    gordonEnvironmentalRadiationImpact2022,
    balSystematicImprovementsTransmon2024,
    biznarovaMitigationInterfacialDielectric2024,
    tuokkolaMethodsAchieveNearmillisecond2025,
    blandMillisecondLifetimesCoherence2025b}, and the plotted qubit-lattice
    processors are taken from
    Refs.~\cite{wuStrongQuantumComputational2021a,
    renExperimentalQuantumAdversarial2022,
    xuDigitalSimulationProjective2023,
    xiangEnhancedQuantumState2024,
    zhangDemonstratingQuantumError2025,
    jinTopologicalPrethermalStrong2025,
    caoExactDecodingQuantum2025,
    chenEfficientImplementationArbitrary2025,
    wangProbingManyBodyBell2025,
    googleQuantumAIQuantumErrorCorrection2025,
    alghadeerLowCrosstalkScalable2025,
    jiangOneTwodimensionalCluster2026,
    huangObservationExactQuantum2026,
    yanCharacterizingManybodyDynamics2026,
    romeiroScalableSingleStepGeneration2026}.  
    Appendix~\ref{app_intro_t1_review}
    and Table~\ref{tab:intro_t1_literature_review} list 
    the specific values and source notes.
    }
    \label{fig:intro_t1_landscape}
\end{figure*}

However, the same progress has not automatically translated 
into multiqubit processors. 
As summarized in Fig.~\ref{fig:intro_t1_landscape},
multiqubit processors based on qubit-coupler lattices often report
typical qubit lifetimes well below those of isolated or weakly connected
qubits
\cite{aruteQuantumSupremacyUsing2019,wuStrongQuantumComputational2021a,
renExperimentalQuantumAdversarial2022,xuDigitalSimulationProjective2023,
xiangEnhancedQuantumState2024,zhangDemonstratingQuantumError2025,
jinTopologicalPrethermalStrong2025,caoExactDecodingQuantum2025,
chenEfficientImplementationArbitrary2025,wangProbingManyBodyBell2025,
googleQuantumAIQuantumErrorCorrection2025,
alghadeerLowCrosstalkScalable2025,jiangOneTwodimensionalCluster2026,
huangObservationExactQuantum2026,yanCharacterizingManybodyDynamics2026,
romeiroScalableSingleStepGeneration2026,
liuPrethermalizationRandomMultipolar2026},
even though the underlying loss mechanisms
are not fundamentally different.  
The distinction is that lattice qubits are embedded
in a much more complex environment,
where loss channels such as dielectric
loss and Purcell decay are more difficult to control.
This observation motivates a ``single-to-more'' viewpoint: 
design rules obtained from isolated qubits in the past 
should be tested again after the qubit is
embedded in the connected environment required by a multiqubit processor.

Here we use finite-element simulation to investigate how surface
participation ratios (SPR)~\cite{wennerSurfaceLossSimulations2011a,wangSurfaceParticipationDielectric2015}
and the resulting surface dielectric loss change when a qubit is embedded
in a flip-chip qubit-coupler lattice.
The results show that higher connectivity can indeed lead to larger surface
loss, providing a possible explanation for why qubit lifetimes in multiqubit
lattices are generally lower than those of isolated qubits.
We introduce a metric, the connectivity-induced surface-loss penalty
$R_\Gamma$, to quantify this effect.
We attribute the connectivity-induced surface-loss change to added edge
fields induced by the coupling claws, field redistribution over the larger
connected metal network, and mode hybridization.
We also investigate the dependence of connectivity-induced surface loss on
geometric design parameters and show that some geometry optimizations used
to reduce single-qubit surface loss can have the opposite effect for qubits
in lattice processors.
In addition, we show that circuit elements 
that are not directly part of the qubit electrodes, 
but are capacitively coupled to the qubit, can also modify
the qubit-mode interface participation ratio.
These results emphasize the importance of evaluating qubit surface loss in
the connected lattice, rather than in an isolated-qubit geometry, when
designing processors based on qubit-coupler lattices.

The paper is organized as follows.
We first identify the mechanism of connectivity-induced surface-loss change
in Sec.~\ref{sec:connectivity_spr}.
In Sec.~\ref{sec:parameter_dependence}, we investigate how this effect
depends on geometric design parameters and propose design guidelines for a
low-loss lattice processor in Sec.~\ref{sec:design_principles}.
Finally, we discuss the implications of our simulation results and other
possible channels of connectivity-induced loss in Sec.~\ref{sec:discussion}.

\section{Connectivity-induced SPR change in qubit-coupler lattices}
\label{sec:connectivity_spr}

Surface dielectric loss contributes substantially to the dissipation of
superconducting qubits.  For assumed interface loss tangents, the
surface-loss-limited quality factor and lifetime of mode $m$ can be estimated
as~\cite{wangSurfaceParticipationDielectric2015}
\begin{equation}
    \frac{1}{Q_m^{\rm surf}}
    = \sum_i p_{m,i}\tan\delta_i,
    \qquad
    T_{1,m}^{\rm surf}=\frac{Q_m^{\rm surf}}{\omega_m},
    \label{eq:surface_loss_model}
\end{equation}
where $p_{m,i}$ is the surface participation ratio (SPR), and
$i\in\{\mathrm{MA},\mathrm{MS},\mathrm{SA}\}$ labels the metal-air,
metal-substrate, and substrate-air interfaces.  In this section, we study how
embedding qubits in qubit-coupler lattices changes the SPR and therefore the
surface-loss-limited coherence time of the qubit modes.

\begin{figure*}[ht]
    \centering
    \includegraphics[width=\textwidth]{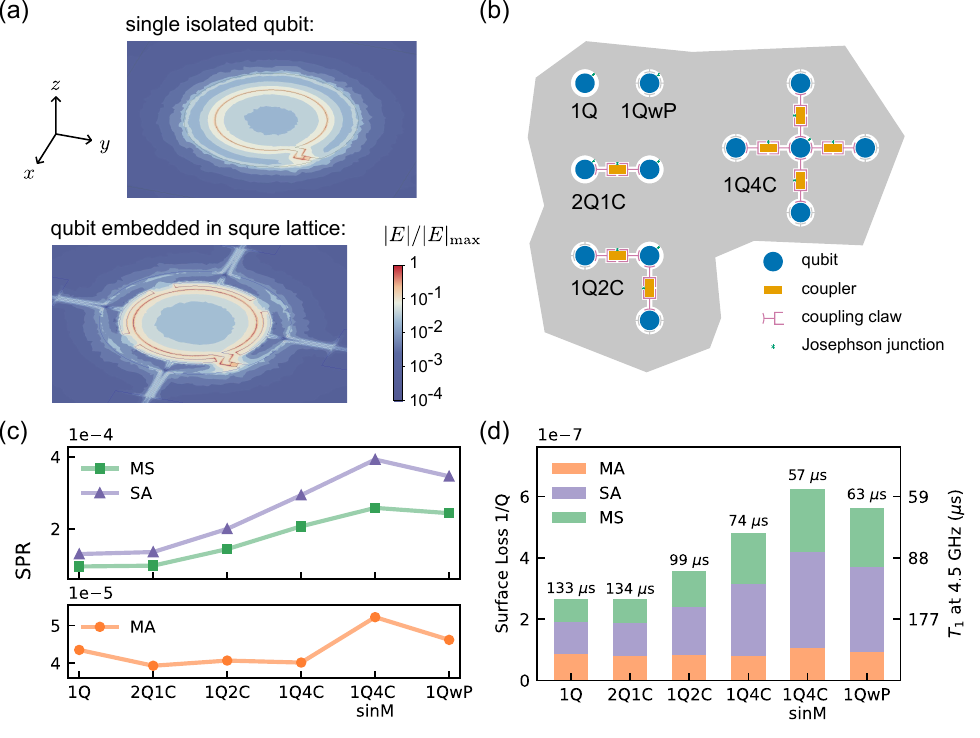}
    \caption{
    Surface-participation simulations of single-qubit and qubit-coupler
    lattice geometries.
    (a) Electric-field distributions of an isolated single qubit and a qubit embedded in a square lattice.
    (b) Schematic geometries of the simulated models.  
    The 1Q and 1QwP models define the isolated-qubit baseline and the passive-claw control, 
    while 2Q1C, 1Q2C, and 1Q4C represent connected qubit-coupler environments.
    (c) Total MA, MS, and SA SPRs for the different geometries.  For the
    connected models 2Q1C, 1Q2C, and 1Q4C, the plotted data correspond to the
    qubit-like dressed mode.  The 1Q4C sinM result is obtained from a
    single-mode solve of the same 1Q4C geometry.  A split linear axis is used
    because the MA participation is about one order of magnitude smaller than
    the MS and SA participations.
    (d) Surface-loss budget $1/Q^{\rm surf}$ calculated from
    Eq.~\eqref{eq:surface_loss_model} using
    $\tan\delta_{\rm MA}=2\times10^{-3}$ and
    $\tan\delta_{\rm MS}=\tan\delta_{\rm SA}=0.8\times10^{-3}$.  The right
    axis gives the corresponding $T_1^{\rm surf}$ at
    $\omega_q/2\pi=4.5\,\mathrm{GHz}$.
    }
    \label{fig:spr_diff_C}
\end{figure*}

The effect of coupler connectivity on qubit SPR is governed by several mechanisms.
As shown in Fig.~\ref{fig:spr_diff_C}(a), couplers introduce additional metal edges, coupling claws, and narrow gaps near the qubit capacitor, all of which can enhance the
electric field at lossy interfaces.  This added-edge effect tends to increase
the SPR and reduce $T_1^{\rm surf}$.
At the same time, the connected metal network changes the capacitance matrix and redistributes the qubit-mode electric field.
In addition, the qubit modes in a square lattice connected with couplers are dressed modes hybridized by bare qubit and coupler modes.
This hybridization will further reshape the field distribution of the qubit-like modes and thus change their SPR.
These effects together contribute to the connectivity-induced SPR variation:
\begin{equation}
    \Delta p_{m,i}
    = \Delta p_{m,i}^{\rm claw}
    + \Delta p_{m,i}^{\rm dis}
    + \Delta p_{m,i}^{\rm hybrid}.
    \label{eq:delta_SPR_decomposition}
\end{equation}
The first term denotes the additional edge and gap participation introduced by coupling claws. The second denotes field redistribution over the larger connected metal network. The third denotes changes caused by hybridization with coupler modes.
The claw contribution generally increases
the SPR, whereas redistribution and hybridization may dilute the qubit-mode
electric field and reduce the SPR.  Consequently, the participation of the dressed-qubit mode need not increase monotonically with the number of attached couplers, and thus quantitative
evaluation is necessary.

\begin{table*}[ht]
    \centering
    \caption{Surface participation ratios and surface-loss metrics for the
    controlled connectivity ladder.
    For the connected models 2Q1C, 1Q2C,
    and 1Q4C, the listed data correspond to the qubit-like dressed mode of
    one qubit connected to one, two, and four couplers, respectively.
    $R_\Gamma$ is the surface-loss rate relative to the isolated single-qubit baseline.
    $R_\Gamma$ and $T_1^{\rm surf}$ are evaluated using the loss tangents in the text.
    More detailed SPR components are listed in Table~\ref{tab:appendix_detailed_participation}.
    }
    \label{tab:connectivity_spr_metrics}
    \begin{tabular}{l l c c c c c}
        \hline\hline
        design & model description
        & $10^4p_{\rm MA}$ & $10^4p_{\rm MS}$ & $10^4p_{\rm SA}$
        & $R_\Gamma$ & $T_1^{\rm surf}$ ($\mu\mathrm{s}$) \\
        \hline
        1Q & isolated single-qubit baseline
        & 0.44 & 0.95 & 1.30 & 1.00 & 132.7 \\
        1QwP & single qubit with four passive coupling claws
        & 0.46 & 2.43 & 3.46 & 2.11 & 62.7 \\
        2Q1C & two qubits connected through a coupler
        & 0.39 & 0.97 & 1.35 & 0.99 & 133.7 \\
        1Q2C & one qubit connected to two couplers
        & 0.41 & 1.43 & 2.00 & 1.33 & 99.5 \\
        1Q4C & one qubit connected to four couplers
        & 0.40 & 2.07 & 2.93 & 1.80 & 73.6 \\
        1Q4C sinM & one qubit connected to four couplers; single-mode solve
        & 0.52 & 2.58 & 3.93 & 2.34 & 56.6 \\
        \hline\hline
    \end{tabular}
\end{table*}

To quantify the connectivity-induced
SPR variation, we apply the two-step finite-element SPR
extraction method of Ref.~\cite{wangSurfaceParticipationDielectric2015} to
qubit-coupler lattice geometries (see Appendix~\ref{app_simlation_method} for details).  
The layouts are based on the flip-chip
qubit-coupler lattice of
Ref.~\cite{liuPrethermalizationRandomMultipolar2026}, 
where Xmon qubits~\cite{barendsCoherentJosephsonQubit2013b} 
are coupled to nearby tunable couplers through 
capacitive coupling pads \cite{liangTunablecouplingArchitecturesCapacitively2023,marxerLongDistanceTransmonCoupler2023}.
The Josephson junctions included in the simulation model are represented as lumped-element inductors, as described in
Appendix~\ref{app_simlation_method}.

The simulated models are shown schematically in
Fig.~\ref{fig:spr_diff_C}(b).  The isolated single-qubit model, denoted 1Q,
defines the reference geometry.  The 1QwP model adds passive coupling claws to
the same qubit and is used to isolate the added-claw contribution
$\Delta p_{m,i}^{\rm claw}$.  The 2Q1C model is the full
qubit-coupler-qubit (QCQ) unit cell, in which two qubits are connected
through one coupler.  The 1Q2C and 1Q4C models are reduced lattice
environments centered on one qubit.  In these two models, junction inductors
are assigned only to the central qubit and to the couplers directly connected
to it.  The outer qubits are retained as capacitive metal structures 
without junction inductors, 
reducing the number of eigenmodes that must be solved.

In the connected models, the bare qubit and coupler frequencies are designed
to be approximately 4.5 and 8 GHz, respectively. 
In the 2Q1C simulation, the
junction inductors of the two qubits are chosen to be slightly different to avoid a near-degeneracy and the associated mixing between the two qubit-like modes.
We solve the active qubit
and coupler modes together in an eigenmode simulation, and
identify the dressed eigenmode near 4.5 GHz as the qubit-like mode whose SPR is reported.  
The 2Q1C, 1Q2C, and 1Q4C simulations therefore test how the
added-claw contribution competes with field redistribution over the connected
metal network and with mode hybridization.  As a control for the hybridization
effect, we also solve the same 1Q4C geometry in a single-mode setting; this
result is labeled 1Q4C sinM.
The lifetimes quoted below are
calculated from Eq.~\eqref{eq:surface_loss_model} using
$\omega_q/2\pi=4.5\,\mathrm{GHz}$,
$\tan\delta_{\rm MA}=2\times10^{-3}$, and
$\tan\delta_{\rm MS}=\tan\delta_{\rm SA}=0.8\times10^{-3}$, which are
reasonable values for aluminum-on-sapphire devices
\cite{wennerSurfaceLossSimulations2011a,wangSurfaceParticipationDielectric2015}.
We also introduce a metric, 
the connectivity-induced surface-loss penalty
$R_{\Gamma,m}=Q_{\mathrm{1Q}}^{\rm surf}/Q_m^{\rm surf}$, 
to quantify the surface-loss rate 
relative to the isolated single-qubit baseline.
The resulting interface participations are plotted in
Fig.~\ref{fig:spr_diff_C}(c), and the corresponding surface-loss budgets are
shown in Fig.~\ref{fig:spr_diff_C}(d).  
The numerical values are summarized in Table~\ref{tab:connectivity_spr_metrics}.

The comparison between 1Q and 1QwP shows the effect of the added-claw term
$\Delta p_{m,i}^{\rm claw}$.  Relative to the isolated qubit, adding four
passive coupling claws increases the MS and SA participations by factors of 2.56 and 2.67, respectively.  The resulting surface-loss rate is 2.11 times larger, reducing the estimated $T_1^{\rm surf}$ from 133 to
63$\,\mu\mathrm{s}$.  This confirms the expected mechanism: coupling claws
and their narrow gaps concentrate the electric field mainly at the substrate
interfaces that dominate the surface-loss model, consistent with the
edge-field sensitivity of surface-participation analyses
\cite{wangSurfaceParticipationDielectric2015,gambettaInvestigatingSurfaceLoss2017}.

The result of 2Q1C shows why connectivity cannot be interpreted as a simple monotonic penalty. 
Its interface participation ratios and estimated $T_1^{\rm surf}$ remain close to those of the isolated-qubit baseline.  
The comparison of 1Q and 1QwP shows that
added claws alone are lossy; the near-baseline 2Q1C
result indicates that field redistribution over the connected metal network
and multimode hybridization can compensate the added-edge penalty for this
particular qubit-like eigenmode.

As the connectivity is increased, the MS and SA participations grow
monotonically with the number of connected couplers.  For the dressed-qubit
mode in 1Q2C, the MS and SA participations increase to 1.51 and 1.54 times
the 1Q baseline, yielding a 1.33-fold surface-loss rate and
$T_1^{\rm surf}=99.5\,\mu\mathrm{s}$.  For the dressed-qubit mode in 1Q4C,
the MS and SA participations further increase to 2.18 and 2.27 times the
1Q baseline, and the total surface-loss rate reaches 1.80 times the baseline,
corresponding to $T_1^{\rm surf}=73.6\,\mu\mathrm{s}$.  
Thus, combining all the effects listed in
Eq.~\eqref{eq:delta_SPR_decomposition}, embedding the qubit in a
multi-coupler lattice leads to larger SPR of the qubit-like modes and
therefore lower $T_1^{\rm surf}$.

Comparing 1Q4C with 1QwP and 1Q4C sinM further separates the
redistribution and hybridization contributions to
Eq.~\eqref{eq:delta_SPR_decomposition}.  First, the comparison between
1Q4C sinM and 1QwP isolates the effect of redistributing the qubit electric
field over the larger connected metal network.  
Relative to 1QwP, the MA, MS, and SA participations
of 1Q4C sinM are changed by factors of 1.13, 1.06, and 1.14, respectively,
and the surface-loss rate increases by a factor of 1.11.  This indicates that
the connected coupler metal network slightly enhances the interface
participations in the single-mode control, rather than simply diluting the
surface fields.
Second, the comparison between the dressed 1Q4C qubit mode and 1Q4C sinM
isolates the effect of mode hybridization.  The dressed 1Q4C qubit mode has
0.77, 0.80, and 0.75 of the MA, MS, and SA participations of 1Q4C sinM,
respectively.  This reduction shows that hybridization with the coupler modes
can dilute the concentrated qubit-mode surface fields.  Finally, comparing
the dressed 1Q4C qubit mode directly with the claw-only 1QwP model gives the
net effect of field redistribution and mode hybridization in the connected
lattice.  The dressed 1Q4C mode has 0.87, 0.85, and 0.85 of the MA, MS, and
SA participations of 1QwP, and its surface-loss rate is reduced to 0.85 of
the 1QwP value,
even though the two models contain the same coupling-claw structure. 
This comparison shows that field redistribution over the connected metal network and mode
hybridization together dilute the qubit-mode electric field
and reduce the resulting SPR.

The design implication is that embedding qubits in a square lattice with
couplers can increase the surface loss of the qubit-like modes, but the effect
is not determined by a simple geometric count of couplers.
Coupling claws and
narrow gaps provide a robust added-edge participation channel, whereas field redistribution over the larger connected metal
network and mode hybridization can either reduce or enhance the SPR of a particular qubit-like eigenmode. 
Low-loss lattice
design should therefore optimize the qubit-coupler coupling claws to suppress
claw-induced SPR increase, while also optimizing the global lattice geometry
to dilute the electric fields or redistribute them away from lossy
interfaces.

\section{Design-parameter dependence of connectivity-induced surface-loss penalty}
\label{sec:parameter_dependence}

\begin{figure*}[h]
    \centering
    \includegraphics[width=6.5in]{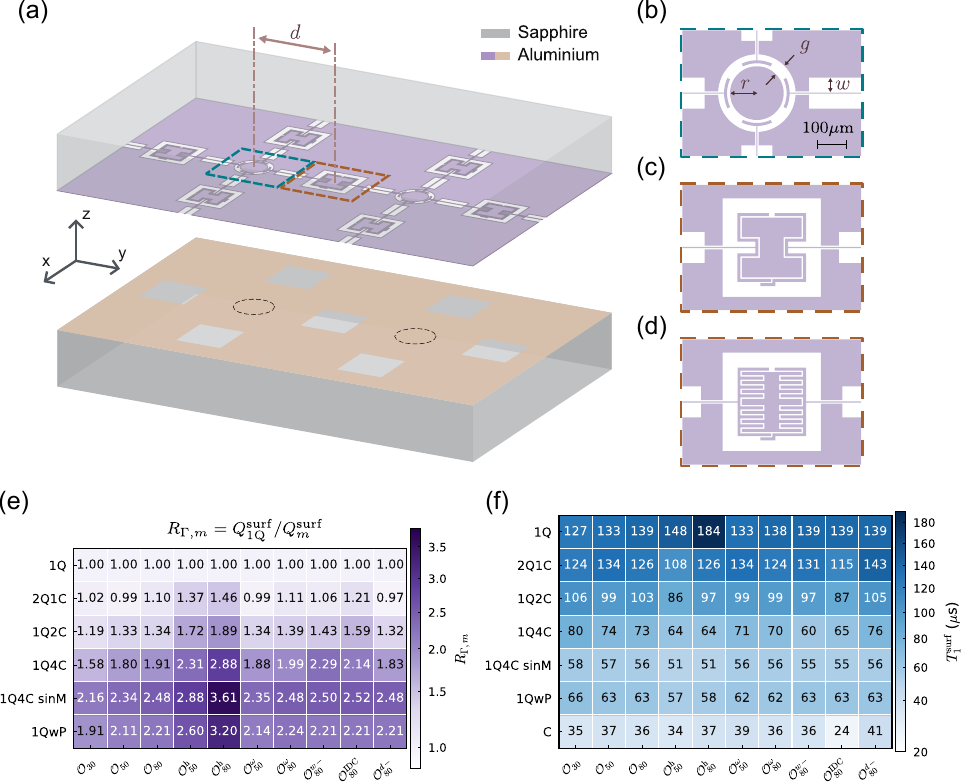}
    \caption{
    Design-parameter dependence of connectivity-induced surface-loss penalty.
    (a) and (b) define the flip-chip layout and local geometry controls:
    the qubit island radius $r$, the qubit-ground gap $g$, the coupling-claw
    gap $w$, and the claw length $d$.  
    The dashed circles on the bottom chip in panel (a) indicate optional
    holes etched in the ground plane below the qubit. 
    (c) and (d) show schematics of the coupler designs; 
    (d) corresponds to design $\mathcal{O}_{80}^{\rm IDC}$, 
    which uses an IDC
    to achieve large capacitive coupling between the claw and the coupler.
    The Josephson junctions in the qubits and couplers are omitted from
    these schematics.
    The resulting $R_\Gamma$ and qubit-mode $T_1^{\rm surf}$ maps are
    summarized in panels (e) and (f), respectively.
    The coupler-mode $T_1^{\rm surf}$ is also shown in the last row of
    panel (f).
    Note that $\mathcal{O}_{80}^{w-}$, $\mathcal{O}_{80}^{\rm IDC}$,
    and $\mathcal{O}_{80}^{d-}$ share the same 1Q and 1QwP models as
    $\mathcal{O}_{80}$.
    }
    \label{fig:design_parameter_sweep}
\end{figure*}

\begin{table*}[h]
    \centering
    \small
    \setlength{\tabcolsep}{4pt}
    \caption{
    Compact notation for the design-parameter sweep.  The baseline family
    $\mathcal{O}_{g}$ varies the qubit-ground gap $g$ at fixed qubit radius
    $r=90\,\mu\mathrm{m}$.  Superscripts denote the controlled geometric or
    simulation change relative to the corresponding baseline design.
    }
    \label{tab:design_label_scheme}
    \begin{tabular}{l l l}
        \hline\hline
        manuscript label & varied control & meaning \\
        \hline
        $\mathcal{O}_{30}$, $\mathcal{O}_{50}$, $\mathcal{O}_{80}$
        & $g$ & qubit-ground gap of 30, 50, or 80 $\mu\mathrm{m}$ \\
        $\mathcal{O}_{50}^{\rm h}$, $\mathcal{O}_{80}^{\rm h}$
        & $h$ & adding opposite-chip hole below the qubit \\
        $\mathcal{O}_{50}^{\omega}$, $\mathcal{O}_{80}^{\omega}$
        & $\omega_q$ & same geometry, shifted qubit-mode frequency \\
        $\mathcal{O}_{80}^{w-}$
        & $w$ & reduced coupling-claw gap \\
        $\mathcal{O}_{80}^{\rm IDC}$
        & $C_{pc}$ & larger claw-coupler capacitance using an IDC \\
        $\mathcal{O}_{80}^{d-}$
        & $d$ & shorter coupling claw \\
        \hline\hline
    \end{tabular}
\end{table*}

\begin{figure*}[tbp]
    \centering
    \includegraphics[width=6.5in]{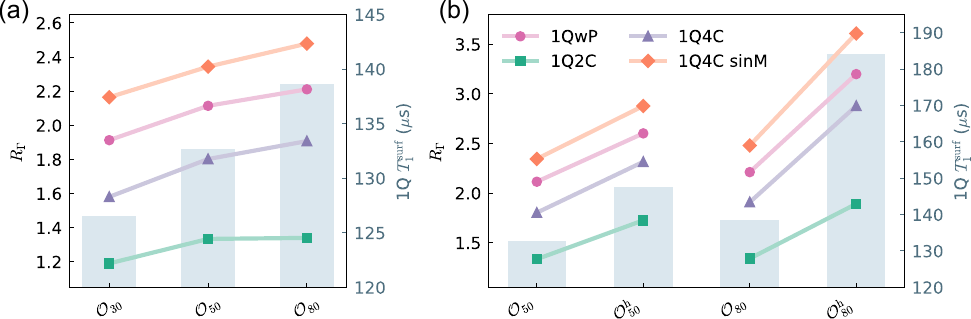}
    \caption{
    Gap and opposite-chip-hole controls for the qubit designs.
    (a) Gap sweep, with
    qubit $R_\Gamma$ shown as markers and curves on the left axis and
    isolated-qubit $T_1^{\rm surf}$ shown as columns on the right axis.
    (b) Opposite-chip-hole comparison, using the same two-axis convention as (a).
    }
    \label{fig:gap_hole_mechanism}
\end{figure*}

The preceding section demonstrates that connectivity changes
the surface participation ratio of qubit modes through added claws, field
redistribution, and mode hybridization, 
and thus leading to connectivity-induced T1 degradation.
Then a natural question is 
how to optimize the qubit-coupler lattice design 
to alleviate this connectivity-induced surface loss.
In this section, we address this question 
by carrying out the same simulation procedure 
on a set of devices with varied qubit-coupler geometries.
As summarized in Fig.~\ref{fig:design_parameter_sweep},  
the devices used in the simulation have a circular qubit island 
with fixed radius $r=90\,\mu\mathrm{m}$, 
while the qubit-ground gap $g$, opposite-chip hole, 
coupling-claw gap $w$, claw-coupler capacitance $C_{pc}$, 
and claw length $d$ are varied.  
Table~\ref{tab:design_label_scheme} defines the compact
notation used to label them below.  
The simulation results are summarized 
in Fig.~\ref{fig:design_parameter_sweep}(e) and (f),
using the same surface-loss model and loss tangents as in
Sec.~\ref{sec:connectivity_spr}.

We first compare the gap sweep
$\mathcal{O}_{30}$, $\mathcal{O}_{50}$, and $\mathcal{O}_{80}$.  
Increasing $g$ lowers the isolated-qubit surface participation 
and improves the single-qubit estimate 
from $T_1^{\rm surf}=127\,\mu\mathrm{s}$ to
$133\,\mu\mathrm{s}$ and then to $139\,\mu\mathrm{s}$, 
as shown in Fig.~\ref{fig:gap_hole_mechanism} (a).  
This trend is consistent with the intuition that 
increasing the spacing between the qubit island and the surrounding ground 
can dilute the localized electromagnetic field in interface-dense regions
~\cite{wangSurfaceParticipationDielectric2015,gambettaInvestigatingSurfaceLoss2017},
and thus reduce the qubit's sensitivity to surface dielectric loss.  
The connectivity penalty $R_\Gamma$, however, 
follows the opposite tendency.
As shown in Fig.~\ref{fig:gap_hole_mechanism} (a), when $g$ increases from 30 to 50 and then to 80 $\mu\mathrm{m}$, 
the resulted $R_\Gamma$ increases from 1.19 to 1.33 and 1.34 for 1Q2C, 
from 1.58 to 1.80 and 1.91 for 1Q4C. 
The same trend appears in the passive-claw model 1QwP, 
where $R_\Gamma$ increases from 1.91 to 2.11 and then to 2.21.  
We interpret this as a competition 
between field dilution and local screening. 
The nearby ground plane can screen the electric field 
between the qubit island and the coupling claws, 
weakening the claw-induced contribution to $R_\Gamma$.  
When the ground plane is moved farther away, 
this screening is weakened, 
the effective qubit-claw field is less suppressed,
and the claw-induced contribution to $R_\Gamma$ becomes larger. 
Thus, a larger qubit-ground gap improves
$T_1^{\rm surf}$ of an isolated qubit, but can also increase the
connectivity-induced penalty $R_\Gamma$ of qubits 
embedded in the qubit-coupler lattice.

Opposite-chip holes show the competition between field dilution 
and ground-plane screening more clearly, 
as shown in Fig.~\ref{fig:gap_hole_mechanism} (b).  
Removing the opposite ground plane below the qubit
reduces the isolated-qubit surface participation 
and improves the estimated single-qubit lifetime.  
At the same time, the connectivity-induced penalties increase.  
For example, for the device with $g=80\,\mu\mathrm{m}$, 
adding the opposite-chip hole increases $R_\Gamma$ 
from 1.34 to 1.89 for 1Q2C,
from 1.91 to 2.88 for 1Q4C. 
The 1QwP control also increases
strongly, from 2.21 to 3.20.  
These trends indicate that the hole improves the
single-qubit $T_1^{\rm surf}$ by diluting the field 
between the qubit and the opposite ground plane,
but weakens the screening of the coupling-claw region 
by the nearby ground plane.  
The weakened screening enhances the claw contribution to SPR 
and therefore increases the connectivity-induced penalty.  

We notice that the hole designs also shift 
the solved qubit frequencies upward.
To check whether the $T_1^{\rm surf}$ and $R_\Gamma$ changes 
are actually a consequence of this frequency shift,
we perform frequency-control simulations in Appendix~\ref{app_frequency_control}.
The results show that this frequency shift produces much smaller changes 
in $R_\Gamma$ than the hole geometry itself.  
The observed $T_1^{\rm surf}$ and penalty changes are therefore
primarily geometric rather than a consequence of the shifted eigenmode
frequency.

\begin{figure}[h]
    \centering
    \includegraphics[width=3.37in]{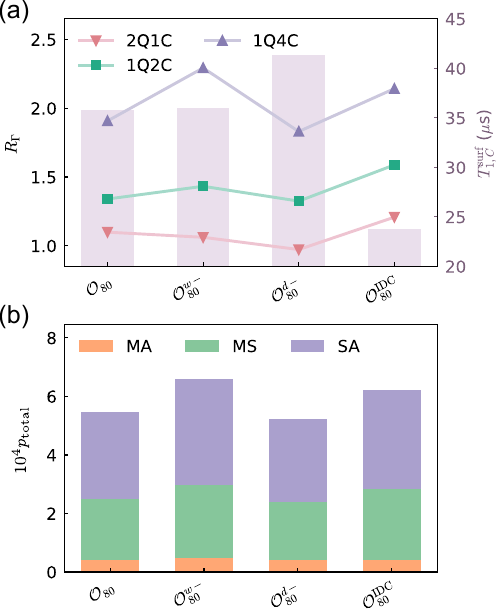}
    \caption{
    Coupling-claw controls around $\mathcal{O}_{80}$.
    (a) Qubit-mode connectivity-induced penalties $R_\Gamma$ 
    for the investigated designs: reduced claw
    gap, shorter claw, and IDC design, 
    with the coupler-mode surface-loss-limited
    lifetime shown as columns on the right axis.  
    (b) MA/MS/SA participation
    decomposition of the qubit mode in 1Q4C for the same designs.  
    }
    \label{fig:claw_coupling_controls}
\end{figure}

So far, we have mainly varied the qubit-side geometry.
The resulting changes in the qubit-mode $T_1^{\rm surf}$ and
$R_\Gamma$ are therefore expected, since these parameters directly
modify the qubit geometry.
We next ask a subtler question: can the geometry of elements that are
not part of the qubit itself, but are connected to it, also affect the
qubit-mode $T_1^{\rm surf}$ and $R_\Gamma$?
To address this question, we vary the geometry of the coupling claw,
which is capacitively coupled to the qubit but is not part of the qubit
itself.
Specifically, starting from $\mathcal{O}_{80}$, we generate several
variants by modifying the coupling-claw geometry
(see Appendix~\ref{app_qcq_capacitance} for details) and compare their
results in Fig.~\ref{fig:claw_coupling_controls}.

We first vary the coupling-claw gap $w$ and length $d$, as labeled in
Fig.~\ref{fig:design_parameter_sweep}(a) and (b).
Compared with $\mathcal{O}_{80}$, $\mathcal{O}_{80}^{w-}$ increases
the qubit-mode $R_\Gamma$ in 1Q2C and 1Q4C, as shown in
Fig.~\ref{fig:claw_coupling_controls}(a).
We then shorten the coupling claw by 100 $\mu$m, yielding
$\mathcal{O}_{80}^{d-}$.
Compared with $\mathcal{O}_{80}$, this variation slightly reduces the
qubit-mode $R_\Gamma$ and also increases the coupler-mode
$T_1^{\rm surf}$.
The 1Q4C SPR decomposition in Fig.~\ref{fig:claw_coupling_controls}(b)
shows that the observed changes are primarily associated with the MS
and SA participations.
Together, these results support the physical picture that part of the
qubit-mode energy is stored in the coupling-claw region.
In this picture, reducing the physical extent $d$ of this region, or
reducing its electric-field concentration by increasing the
claw--ground spacing $w$, weakens the additional MS/SA surface-loss
channel seen by the qubit modes, thereby reducing the
connectivity-induced penalty.

Finally, increasing the claw--coupler capacitance $C_{pc}$ with an
interdigitated capacitor (IDC) yields $\mathcal{O}_{80}^{\rm IDC}$ and
produces a pronounced change.
As shown in Fig.~\ref{fig:claw_coupling_controls}, 
the qubit-mode $R_\Gamma$ rise to 1.21, 1.59,
and 2.14 for 2Q1C, 1Q2C, and 1Q4C, respectively, 
while the coupler-mode $T_{1}^{\rm surf}$ 
decreases to $24\,\mu\mathrm{s}$.  
The $T_{1}^{\rm surf}$ degradation of the coupler mode is expected, 
because IDC structures introduce long narrow gaps around the coupler, 
which leads to larger interface participation.
And about the increased qubit-mode $R_\Gamma$, 
this phenomenon actually confirms the physical picture we conveyed 
in Sec.~\ref{sec:connectivity_spr}:
in the qubit-coupler lattice, the field of the qubit mode 
is redistributed over the connected metal network,
and part of the qubit-mode electric field is also distributed 
in the coupling-claw region.
Therefore, changing the claw-coupler geometry, 
which is not part of the qubit itself,
can also modify the qubit-mode interface participation ratio 
and thus the resulting interface dielectric loss.
Thus, increasing $C_{pc}$ not only degrades the coupler mode but also
increases the surface loss of the qubit modes.
The above investigation of varying the coupling-claw geometry 
show that the qubit $T_1^{\rm surf}$ is not controlled only by 
the local qubit geometry:
the geometry of nonlocal circuit elements coupled to the qubit can also modify
the qubit-mode interface participation ratio and thus the resulting interface
dielectric loss.

The solved coupler-mode frequencies in these SPR simulations are close to the
corresponding coupling-off frequencies $f_{c,\rm off}$ listed in
Appendix~\ref{app_qcq_capacitance}.  
Thus, the above surface-loss estimates 
mainly characterize the near-idle regime of the qubit-coupler lattice, 
rather than the full range of coupler biases encountered during a gate.

\section{Design principles for low-loss qubit-coupler lattices}
\label{sec:design_principles}

In Sec.~\ref{sec:parameter_dependence}, 
we investigate how the qubit-mode $R_\Gamma$ 
and coupler-mode $T_{1}^{\rm surf}$
depend on the geometry design of qubit-coupler lattice,
including the qubit-ground gap, opposite-chip hole, and coupling-claw geometry.
Here in this section, we summarize the design principles for low-loss qubit-coupler lattices
based on the simulation results in Sec.~\ref{sec:parameter_dependence}.

First, a design with a better single-qubit $T_1^{\rm surf}$ 
can still suffer an even larger connectivity-induced penalty $R_\Gamma$ 
after embedding in the qubit-coupler lattice.
The qubit-ground gap and opposite-chip hole examples shown above demonstrate 
this point clearly.
Therefore, when designing a processor with qubit-coupler lattices, 
it is necessary to consider and simulate its performance 
in the connected lattice, like evaluating the qubit-mode $R_\Gamma$,
rather than only focusing on the isolated-qubit $T_1^{\rm surf}$.
And some design principles 
used to obtain a better single-qubit $T_1^{\rm surf}$,
like diluting the qubit electric field 
by increasing the qubit-ground gap or adding an opposite-chip hole,
may lead to a larger connectivity-induced penalty $R_\Gamma$,
and thus are not suitable for the qubit-coupler lattice design.

Second, the geometry of the circuit elements 
that are not directly part of the qubit electrodes,
but are capacitively coupled to it, 
can also modify the qubit-mode interface participation ratio.
As shown in the coupling-claw geometry controls 
in Sec.~\ref{sec:parameter_dependence},
changing the geometry of the coupling claws, 
which are not part of the qubit itself,
can also modify the qubit-mode interface participation ratio 
and thus the resulting interface dielectric loss.
This is because a fraction of the qubit-mode electric field 
is actually distributed in the coupling-claw region.
This message again emphasizes the importance of 
simulating the qubit-coupler lattice rather than a single qubit
when designing a multiqubit processor.
Beyond qubit-mode coherence, the coupler-mode $T_{1}^{\rm surf}$ should also be
evaluated to ensure that the coupler mode is not too lossy, since
coupler coherence can affect two-qubit gate fidelity.  
Apart from surface loss discussed here, parameters
such as coupling strengths are also important metrics for assessing the performance
of a qubit-coupler lattice; this aspect is addressed in
Appendix~\ref{app_qcq_capacitance}.

Third, the specific distribution and decomposition of 
the interface participation ratio should guide fabrication priorities.  
The detailed diagnostics in
Fig.~\ref{fig:simulation_detail_appendix} show that, 
for the simulated flip-chip transmon geometries, 
the MS and SA participations are usually larger than MA,
consistent with previous surface-participation analyses 
on single-chip devices 
\cite{wennerSurfaceLossSimulations2011a,wangSurfaceParticipationDielectric2015}.
In addition, the top-bottom decomposition in
Fig.~\ref{fig:top_bottom_spr_appendix} shows that the MS and SA
participations are strongly localized on the top chip, while the auxiliary
substrate-energy partition in Table~\ref{tab:substrate_energy_partition}
shows that the top substrate also 
stores the largest fraction of bulk electric energy.
In other words, for a flip-chip transmon lattice,
the chip carrying the qubit electrodes accounts for 
the majority of the qubit-mode interface participation.
Therefore, improving the quality and cleanliness of 
the top-chip metal-substrate and substrate-air interfaces 
is especially important for suppressing the dielectric loss 
of both the qubits and couplers.

\section{Discussion}
\label{sec:discussion}

Our results identify connectivity-induced SPR increase 
as a geometry-dependent loss channel 
that is absent in isolated-qubit optimization
but naturally appears when qubits are embedded in a qubit-coupler lattice.
This mechanism provides a plausible contribution to
the persistent gap between the longest reported lifetimes of isolated
superconducting qubits and the typical lifetimes of qubits operated in large
processor lattices.  
It also shows that low-loss lattice design requires metrics that explicitly
quantify the cost of connectivity.  The connectivity-induced penalty $R_\Gamma$ 
provides one such metric by comparing the surface-loss rate of 
a dressed qubit mode embedded in the qubit-coupler lattice 
with the corresponding isolated-qubit reference, 
and should be considered 
when designing multiqubit processors based on qubit-coupler lattices.

The estimated surface-loss-limited lifetimes reported here
should be interpreted as comparative surface-loss estimates 
rather than absolute predictions for a fabricated device.
We assume $\tan\delta_{\rm MA}=2\times10^{-3}$, and
$\tan\delta_{\rm MS}=\tan\delta_{\rm SA}=0.8\times10^{-3}$
in our simulation.
The actual loss tangents of these interfaces may depend on
the specific fabrication process and materials used,
and can be extracted 
using experimental characterization 
~\cite{wangSurfaceParticipationDielectric2015,hedrickQuantifyingSurfaceLosses2026}.
Our simulation mainly provides guidelines of 
reducing dielectric losses from the geometry design perspective.
Optimization of the fabrication process and materials is 
another important direction
for reducing dielectric losses in superconducting circuits.
For example, using materials, like 
Ta~\cite{placeNewMaterialPlatform2021,wangPracticalQuantumComputers2022,
ganjamSurpassingMillisecondCoherence2024,blandMillisecondLifetimesCoherence2025b}, 
Nb~\cite{tuokkolaMethodsAchieveNearmillisecond2025}
or TiN~\cite{dengTitaniumNitrideFilm2023,yanCharacterizingManybodyDynamics2026}, 
that are compatible with HF cleaning processes,
may help to improve the quality of the interfaces.
And still using Al but with methods to improve its interface quality
~\cite{biznarovaMitigationInterfacialDielectric2024,mahuliImprovingLifetimeAluminumBased2025}
also shows promise for improving the qubit $T_1$.

Apart from the surface dielectric loss, 
other loss channels may also account for the reduced $T_1$ of 
qubits embedded in large lattices.
For example, to realize manipulation of a multiqubit processor,
large numbers of control and readout lines are required,
and these waveguides may be strayly coupled to the qubit and coupler modes,
leading to additional Purcell decay.
In addition, packaging multiqubit superconducting circuits often
require relatively large enclosures, 
for which suppressing spurious modes and
seam loss remains an important engineering challenge~\cite{huangMicrowavePackageDesign2021}.

\section*{Acknowledgements}

This work was supported by the
National Natural Science Foundation of China (Grants
No. 92265207, No. T2121001,  No. T2322030, No. 12122504, No. 12274142,
No. 92365206, No. 12104055, No. 12475017, No. U25A6009), 
the Natural Science Foundation of Guangdong Province 
(Grant No. 2024A1515010398),
the Quantum Science and Technology-National 
Science and Technology Major Project
(Grants No. 2021ZD0301800, No. 2021ZD0301802), 
the Beijing Nova Program (No. 20220484121),
the MOST program (Grant No. 2025YFE0217600). 

\section*{Data availability}
The data are available from the authors upon reasonable request.

\appendix	
\section{Summary of reported qubit $T_1$ values in recent literature}
\label{app_intro_t1_review}

Table~\ref{tab:intro_t1_literature_review} lists the literature-reported
energy-relaxation values plotted in Fig.~\ref{fig:intro_t1_landscape}.
In isolated-qubit chips, several recent devices report best
or maximum $T_1$ values above $500\,\mu{\rm s}$, and one entry reaches the
millisecond range.  By contrast, the typical $T_1$ values reported for the
qubit-lattice processors collected here are mostly in the range of several
tens to roughly $150\,\mu{\rm s}$.  

\begin{table*}[t]
    \centering
    \footnotesize
    \setlength{\tabcolsep}{3.2pt}
    \caption{
    Detailed information of the literature values 
    used in Fig.~\ref{fig:intro_t1_landscape}.
    Typical values denote the reported mean, median, or average $T_1$ when such
    a statistic is available.  Best values denote reported maximum or best
    measured $T_1$.  The plot-ID column maps entries to the labels in
    Fig.~\ref{fig:intro_t1_landscape}; a dash indicates that the entry is
    retained for reference but not plotted in Fig.~\ref{fig:intro_t1_landscape}.
    The qubit-number column is used only for qubit-lattice processors.
    }
    \label{tab:intro_t1_literature_review}
    \begin{tabular}{@{}p{0.11\textwidth} c c p{0.12\textwidth} c c p{0.25\textwidth}@{}}
        \hline\hline
        category & plot ID & year & qubit number & typical $T_1$ ($\mu$s)
        & best $T_1$ ($\mu$s) & context and source \\
        \hline
        isolated qubit & I1 & 2021 & -- & 230 & 360
        & Place et al., Princeton~\cite{placeNewMaterialPlatform2021} \\
        isolated qubit & I2 & 2022 & -- & -- & 503
        & Wang et al., BAQIS~\cite{wangPracticalQuantumComputers2022} \\
        isolated qubit & I3 & 2022 & -- & 375 & 600
        & Gordon et al., IBM~\cite{gordonEnvironmentalRadiationImpact2022} \\
        isolated qubit & I4 & 2024 & -- & 323 & 586
        & Bal et al., SQMS~\cite{balSystematicImprovementsTransmon2024} \\
        isolated qubit & I5 & 2024 & -- & 270 & 501
        & Biznarova et al., Chalmers~\cite{biznarovaMitigationInterfacialDielectric2024} \\
        isolated qubit & I6 & 2025 & -- & 425 & 666
        & Tuokkola et al., Aalto~\cite{tuokkolaMethodsAchieveNearmillisecond2025} \\
        isolated qubit & I7 & 2025 & -- & 450 & 1680
        & Bland et al., Princeton~\cite{blandMillisecondLifetimesCoherence2025b} \\
        processor & -- & 2019 & 53 & 16.0 & --
        & Sycamore, Google~\cite{aruteQuantumSupremacyUsing2019} \\
        processor & P1 & 2021 & 66 & 30.6 & --
        & Zuchongzhi, USTC~\cite{wuStrongQuantumComputational2021a} \\
        processor & P2 & 2022 & 36 (10 used) & 150 & --
        & Ren et al., ZJU~\cite{renExperimentalQuantumAdversarial2022} \\
        processor & P3 & 2023 & 68 & 109.8 & --
        & Xu et al., ZJU~\cite{xuDigitalSimulationProjective2023} \\
        processor & P4 & 2024 & 36 & 141.5 & --
        & Xiang et al., ZJU~\cite{xiangEnhancedQuantumState2024} \\
        processor & P5 & 2025 & 121 & 126.1 & --
        & Zhang et al., ZJU~\cite{zhangDemonstratingQuantumError2025} \\
        processor & P6 & 2025 & 125 (100 used) & 70.3 & --
        & Jin et al., ZJU~\cite{jinTopologicalPrethermalStrong2025} \\
        processor & P7 & 2025 & 72 & 77.8 & --
        & Cao et al., BAQIS~\cite{caoExactDecodingQuantum2025} \\
        processor & P8 & 2025 & 12 used & 70.3 & --
        & Chen et al., BAQIS~\cite{chenEfficientImplementationArbitrary2025} \\
        processor & P9 & 2025 & 73 & 81.4 & --
        & Wang et al., ZJU~\cite{wangProbingManyBodyBell2025} \\
        processor & P10 & 2025 & 72 & 73 & 113
        & Willow-72, Google~\cite{googleQuantumAIQuantumErrorCorrection2025} \\
        processor & P11 & 2025 & 105 & 68 & 119
        & Willow-105, Google~\cite{googleQuantumAIQuantumErrorCorrection2025} \\
        processor & P12 & 2025 & 16 & 71 & 126
        & Alghadeer et al., Oxford~\cite{alghadeerLowCrosstalkScalable2025} \\
        processor & P13 & 2026 & 105 & 49.7 & --
        & Zuchongzhi 3.1, USTC~\cite{jiangOneTwodimensionalCluster2026} \\
        processor & P14 & 2026 & 66 (56 used) & 54.3 & --
        & Huang et al., SUSTech~\cite{huangObservationExactQuantum2026} \\
        processor & P15 & 2026 & 16 & 75.6 & 96
        & Yan et al., RIKEN~\cite{yanCharacterizingManybodyDynamics2026} \\
        processor & P16 & 2026 & 17 (10 used) & 86.4 & --
        & Romeiro et al., TUM~\cite{romeiroScalableSingleStepGeneration2026} \\
        processor & -- & 2026 & 78 & 26.4 & --
        & Chuang-tzu 2.0, IOP~\cite{liuPrethermalizationRandomMultipolar2026} \\
        \hline\hline
    \end{tabular}
\end{table*}

\section{Simulation methods for surface participation ratios}
\label{app_simlation_method}

In this appendix, we describe the simulation procedure used to calculate
surface participation ratios. Our approach follows the two-step
finite-element method introduced in Ref.~\cite{wangSurfaceParticipationDielectric2015},
which combines a coarse three-dimensional simulation of the full system with a
fine two-dimensional cross-sectional simulation of the edge regions. In this
method, each surface dielectric layer is divided into an interior region
($x>x_0$) and a perimeter region ($0<x<x_0$), where $x$ is the distance from
the edge and $x_0$ defines the boundary between the two regions.

The surface participation ratio of the interior region, $p_{i,\mathrm{int}}$,
is obtained from the coarse 3D simulation. To evaluate the energy stored in the
perimeter region, we further divide this region into two halves. The energy in
the inner half ($x_0/2<x<x_0$) can also be extracted from the 3D simulation,
whereas the energy in the outer half ($0<x<x_0/2$) does not converge reliably.
Following Ref.~\cite{wangSurfaceParticipationDielectric2015}, a constant
scaling factor $F_i$ is introduced to obtain the total energy in the perimeter
region from the energy in the inner half:
\begin{equation}
p_{i,\mathrm{peri}}=F_i\,p_{i,\mathrm{periH}},
\quad i=\mathrm{ma}, \mathrm{ms}, \mathrm{sa},
\label{eq:E_peri}
\end{equation}
where $p_{i,\mathrm{peri}}$ and $p_{i,\mathrm{periH}}$ are the participation
ratios in the full perimeter region and in the inner half of the perimeter
region, respectively. The scaling factor $F_i$ depends on the chosen value of
$x_0$ and can be calculated using a local 2D cross-sectional simulation of the
edge region. Here we describe how $p_{i,\mathrm{int}}$ and
$p_{i,\mathrm{periH}}$ are obtained from the global coarse 3D simulation; the
calculation of the scaling factor is presented in Appendix~\ref{app_scalingF}.

In the global 3D simulation, the metal films are modeled as two-dimensional
sheets, and the Josephson junctions are represented by lumped-element inductors
with inductance $L_j$, while the detailed structure of the junction leads is
neglected~\cite{minevEnergyparticipationQuantizationJosephson2021b}. For a
flip-chip device, galvanic contact between the ground planes of the top and
carrier chips is modeled by adding arrays of metal bumps, representing the
indium bumps commonly used in flip-chip fabrication. The field distributions
of the qubit modes are obtained from eigenmode simulations. The energies stored
in the MA, MS, and SA surface layers are then calculated from the surface
fields~\cite{wennerSurfaceLossSimulations2011a}.

Specifically, for a thin dielectric layer with thickness $t_i$ and dielectric
constant $\epsilon_i$, the surface participation ratio is given by
($i=\mathrm{ma}, \mathrm{ms}, \mathrm{sa}$)~\cite{wennerSurfaceLossSimulations2011a,martinisSurfaceLossCalculations2022}
\begin{equation}
p_i=\frac{\epsilon_0\epsilon_i}{2W}t_i\int dA\left| E_i \right|^2,
\label{eq:define_SPR}
\end{equation}
where the volume integral over the thin dielectric layer is replaced by a
surface integral $dA$ of the effective surface electric field $E_i$, and $W$ is
the total capacitive energy of the qubit mode~\cite{minevEnergyparticipationQuantizationJosephson2021b}.
The effective electric field $E_i$ within the thin dielectric layer is
determined from the electric-field boundary conditions,
\begin{equation}
E_{i,||}=E_{\mathrm{env},||},
\label{eq:E_boundary_parallel}
\end{equation}
\begin{equation}
\epsilon_iE_{i,\bot}=\epsilon_{\mathrm{env}}E_{\mathrm{env},\bot},
\label{eq:E_boundary_normal}
\end{equation}
where $E_{i,||}$ and $E_{i,\bot}$ are the components parallel and normal to the
interface, respectively, and $E_{\mathrm{env}}$ is the electric field in the
dielectric material adjacent to the layer.

Because the electric field is perpendicular to a metal surface and the
dielectric layer is thin, we retain only the normal component for the MA and MS
layers, i.e., $E_{\mathrm{ma}}=E_{\mathrm{ma},\bot}$ and
$E_{\mathrm{ms}}=E_{\mathrm{ms},\bot}$. For the SA layer, both
$E_{\mathrm{sa},\bot}$ and $E_{\mathrm{sa},||}$ are included. For the MA and
SA interfaces, the adjacent dielectric material is air (vacuum), corresponding
to $\epsilon_{\mathrm{env}}=1$. For the MS interface, the adjacent dielectric
material is the device substrate, corresponding to
$\epsilon_{\mathrm{env}}=\epsilon_{\mathrm{s}}$, where $\epsilon_{\mathrm{s}}$
is the dielectric constant of the substrate. Combining these relations, the MA,
MS, and SA surface participation ratios can be written as~\cite{wennerSurfaceLossSimulations2011a}
\begin{align}
\frac{p_{\mathrm{ma}}W}{t_{\mathrm{ma}}}
&= \frac{1}{2}\epsilon_0\epsilon_{\mathrm{ma}}\int dA
\left| E_{\mathrm{ma}} \right|^2 \notag\\
&= \frac{1}{2}\epsilon_0\epsilon_{\mathrm{ma}}\int dA
\left| E_{\mathrm{a},\bot}/\epsilon_{\mathrm{ma}} \right|^2 \notag\\
&= \epsilon_{\mathrm{ma}}^{-1}
\left[ \frac{1}{2}\epsilon_0\int dA
\left| E_{\mathrm{a},\bot} \right|^2 \right],
\label{eq:p_ma}
\end{align}

\begin{align}
\frac{p_{\mathrm{ms}}W}{t_{\mathrm{ms}}}
&= \frac{1}{2}\epsilon_{\mathrm{ms}}\epsilon_0\int dA
\left| E_{\mathrm{ms}} \right|^2 \notag\\
&= \frac{1}{2}\epsilon_{\mathrm{ms}}\epsilon_0\int dA
\left| \epsilon_{\mathrm{s}}E_{\mathrm{s},\bot}/\epsilon_{\mathrm{ms}}
\right|^2 \notag\\
&= \left( \frac{\epsilon_{\mathrm{s}}^{2}}{\epsilon_{\mathrm{ms}}} \right)
\left[ \frac{1}{2}\epsilon_0\int dA
\left| E_{\mathrm{s},\bot} \right|^2 \right],
\label{eq:p_ms}
\end{align}

\begin{align}
\frac{p_{\mathrm{sa}}W}{t_{\mathrm{sa}}}
&= \frac{1}{2}\epsilon_{\mathrm{sa}}\epsilon_0\int dA
\left| E_{\mathrm{sa}} \right|^2 \notag\\
&= \frac{1}{2}\epsilon_{\mathrm{sa}}\epsilon_0\int dA
\left( \left| E_{\mathrm{sa},\bot} \right|^2
+\left| E_{\mathrm{sa},||} \right|^2 \right) \notag\\
&= \frac{1}{2}\epsilon_{\mathrm{sa}}\epsilon_0\int dA
\left( \left| E_{\mathrm{a},\bot}/\epsilon_{\mathrm{sa}} \right|^2
+\left| E_{\mathrm{a},||} \right|^2 \right) \notag\\
&= \epsilon_{\mathrm{sa}}^{-1}
\left[ \frac{1}{2}\epsilon_0\int dA
\left| E_{\mathrm{a},\bot} \right|^2 \right] \notag\\
&\quad + \epsilon_{\mathrm{sa}}
\left[ \frac{1}{2}\epsilon_0\int dA
\left| E_{\mathrm{a},||} \right|^2 \right],
\label{eq:p_sa}
\end{align}
where $E_{\mathrm{a}}$ and $E_{\mathrm{s}}$ are the electric fields in air and
in the substrate outside the interface, respectively, and $\epsilon_0$ is the
vacuum permittivity.

Using Eqs.~\eqref{eq:p_ma}--\eqref{eq:p_sa}, the surface participation ratios
of the interfaces can be calculated from the electric fields outside the thin
dielectric layers. Therefore, the thin dielectric layers do not need to be
explicitly included in the simulation model~\cite{wennerSurfaceLossSimulations2011a,dialBulkSurfaceLoss2016a,gambettaInvestigatingSurfaceLoss2017}.
This treatment improves the efficiency of the finite-element simulation,
because explicitly meshing nanometer-scale dielectric layers at the interfaces
is computationally expensive, especially in a 3D model.

In summary, $p_{i,\mathrm{int}}$ and $p_{i,\mathrm{periH}}$ are obtained from
the coarse 3D finite-element simulation using
Eqs.~\eqref{eq:p_ma}--\eqref{eq:p_sa}. Together with the scaling factor $F_i$
extracted from a fine 2D cross-sectional simulation of the edge region
(Appendix~\ref{app_scalingF}), the perimeter contribution is calculated as
$p_{i,\mathrm{peri}}=F_i\,p_{i,\mathrm{periH}}$. The total surface
participation ratio is then
$p_{i,\mathrm{total}}=p_{i,\mathrm{int}}+p_{i,\mathrm{peri}}$.

\begin{table*}[ht]
    \centering
    \caption{
    Detailed energy-participation components for the simulated models in Sec.~\ref{sec:connectivity_spr}.  
    Surface participation ratios are listed as $10^4p$ and are calculated using the method and convention described in Appendices~\ref{app_simlation_method} and \ref{app_scalingF}.  
    }
    \label{tab:appendix_detailed_participation}
    \begin{tabular}{l c c c c c c}
        \hline\hline
        metric & 1Q & 1QwP & 2Q1C & 1Q2C & 1Q4C & 1Q4C sinM \\
        \hline
        $10^4 p_{\rm MA,total}$  & 0.435 & 0.462 & 0.393 & 0.407 & 0.401 & 0.522 \\
        $10^4 p_{\rm MA,int}$    & 0.283 & 0.236 & 0.268 & 0.258 & 0.237 & 0.243 \\
        $10^4 p_{\rm MA,peri}$   & 0.152 & 0.225 & 0.125 & 0.148 & 0.165 & 0.279 \\
        $10^4 p_{\rm MA,top}$    & 0.284 & 0.344 & 0.251 & 0.272 & 0.282 & 0.400 \\
        $10^4 p_{\rm MA,bottom}$ & 0.151 & 0.118 & 0.142 & 0.134 & 0.119 & 0.122 \\
        \hline
        $10^4 p_{\rm MS,total}$  & 0.950 & 2.433 & 0.973 & 1.433 & 2.068 & 2.580 \\
        $10^4 p_{\rm MS,int}$    & 0.512 & 0.967 & 0.550 & 0.704 & 0.918 & 0.932 \\
        $10^4 p_{\rm MS,peri}$   & 0.438 & 1.466 & 0.423 & 0.729 & 1.149 & 1.648 \\
        $10^4 p_{\rm MS,top}$    & 0.950 & 2.433 & 0.973 & 1.433 & 2.067 & 2.580 \\
        $10^4 p_{\rm MS,bottom}$ & 0.000 & 0.000 & 0.000 & 0.000 & 0.000 & 0.000 \\
        \hline
        $10^4 p_{\rm SA,total}$  & 1.296 & 3.459 & 1.351 & 1.995 & 2.935 & 3.926 \\
        $10^4 p_{\rm SA,int}$    & 0.687 & 1.208 & 0.808 & 0.950 & 1.216 & 1.074 \\
        $10^4 p_{\rm SA,peri}$   & 0.609 & 2.251 & 0.542 & 1.045 & 1.718 & 2.851 \\
        $10^4 p_{\rm SA,top}$    & 1.296 & 3.459 & 1.351 & 1.995 & 2.934 & 3.925 \\
        $10^4 p_{\rm SA,bottom}$ & 0.000 & 0.000 & 0.000 & 0.000 & 0.001 & 0.001 \\
        \hline\hline
    \end{tabular}
\end{table*}

\section{Scaling factors of flip-chip devices}
\label{app_scalingF}

As discussed in Appendix~\ref{app_simlation_method}, we apply the two-step
simulation method of Ref.~\cite{wangSurfaceParticipationDielectric2015} to a
flip-chip qubit geometry. The global 3D simulation provides the chip-scale
field distribution, from which we calculate $p_{i,int}$ and $p_{i,periH}$,
whereas the 2D cross-section simulation provides multiplicative edge
corrections for the perimeter regions of the interfaces through scaling factors
$F_i$. In this appendix, we describe the simulation of these scaling factors
for the flip-chip geometry.

\begin{figure*}[t]
    \centering
    \includegraphics[width=0.96\textwidth]{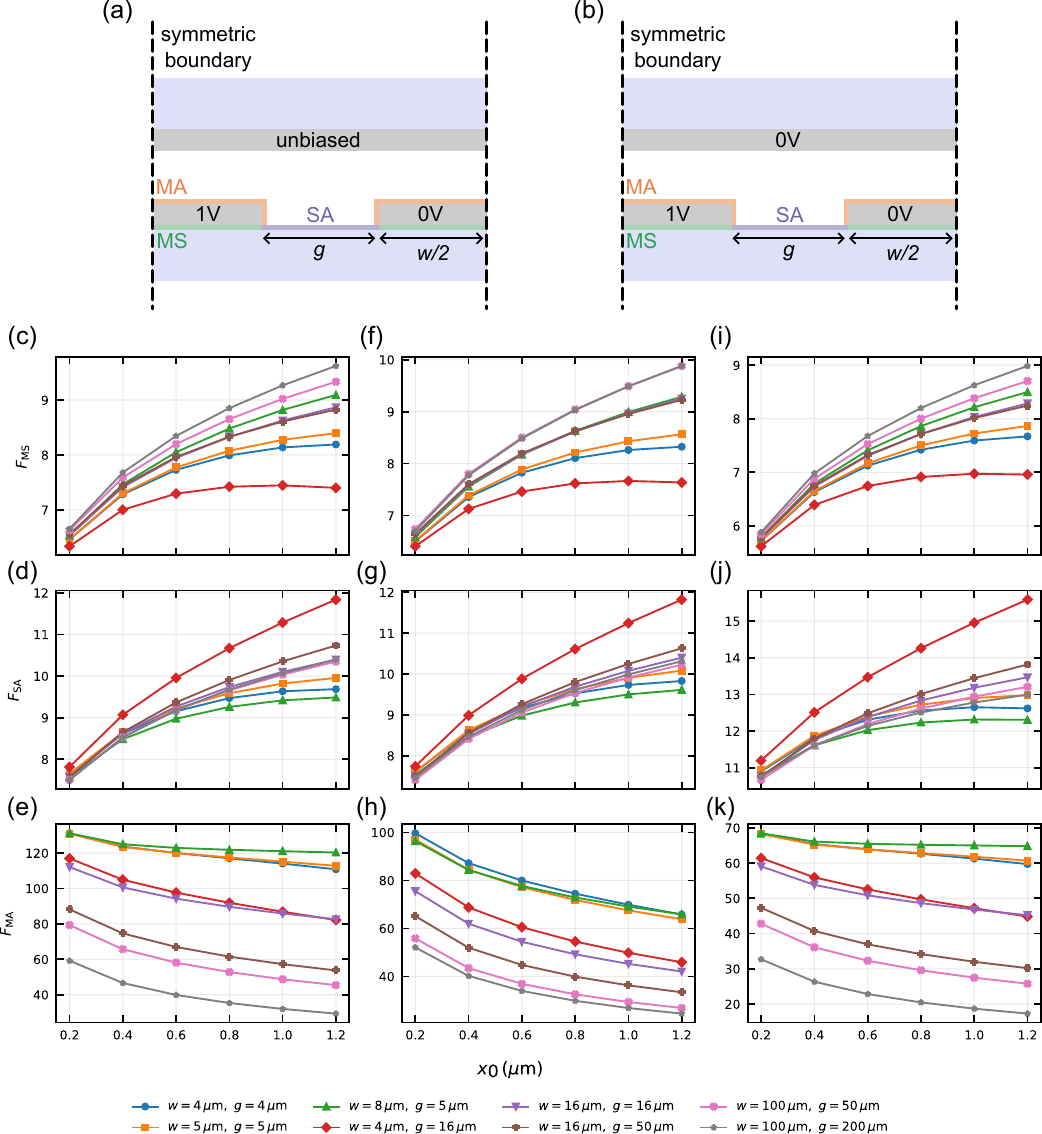}
    \caption{
    Simulation of scaling factors in a flip-chip device.
    Panels (a) and (b) show schematics of the 2D cross-section simulation
    model. The thicknesses of the metal pads, substrates, and all dielectric
    layers are set to 0.1 $\mathrm{\mu m}$, 430 $\mathrm{\mu m}$, and 3 $nm$,
    respectively. The distance between the top and bottom chips is set to
    8 $\mathrm{\mu m}$. The two side boundaries are assigned symmetric boundary
    conditions to represent an interdigitated-capacitor-style device with
    conductor width $w$ and gap width $g$. Panel (a) corresponds to the
    unbiased opposite-plane model, in which the left metal is biased at
    $1\,\mathrm{V}$, the right metal is held at $0\,\mathrm{V}$, and the
    opposite ground plane is left unbiased. Panel (b) corresponds to the
    grounded opposite-plane model, in which the right metal and the opposite
    ground plane are tied to $0\,\mathrm{V}$. Panels (c)--(e), (f)--(h), and
    (i)--(k) show the simulated scaling factors as functions of $x_0$ for
    selected $(w,g)$ values. The results in (c)--(e) and (i)--(k) are obtained
    using the unbiased opposite-plane model in (a) with surface-layer
    permittivities $\epsilon_{\mathrm{layer}}=10$ and
    $\epsilon_{\mathrm{layer}}=4$, respectively. The results in (f)--(h) are
    obtained using the grounded opposite-plane model in (b) with
    $\epsilon_{\mathrm{layer}}=10$.
    }
    \label{fig:scaling_factor_fc_models}
\end{figure*}

The simulation model is a two-dimensional cross section of the capacitor pads
in a flip-chip environment, as shown in
Fig.~\ref{fig:scaling_factor_fc_models}. It includes the lower-chip substrate
and electrodes, the opposite-chip ground plane, and explicit MA, MS, and SA
dielectric interface layers. For simplicity, the three types of interface
layers are assumed to have the same thickness and dielectric constant, set to
$t=3\,\mathrm{nm}$ and $\epsilon_{\mathrm{layer}}=10$, following values
commonly used in surface-loss estimates
\cite{wennerSurfaceLossSimulations2011a,wangSurfaceParticipationDielectric2015}.
The parameter $x_0$ defines the boundary between the interior and perimeter
regions, as described in Appendix~\ref{app_simlation_method}. Even
electrostatic symmetry boundaries are imposed at the lateral edges so that the
model represents a repeated cell of an interdigitated-capacitor (IDC) geometry
with conductor width $w$ and gap width $g$
\cite{wangSurfaceParticipationDielectric2015}.
From this simulation, we obtain the energies in the inner-half perimeter region
$\mathcal{E}_{i,periH}$ and the full perimeter region $\mathcal{E}_{i,peri}$ of
interface $i$, and define the scaling factor as
$F_i=\mathcal{E}_{i,peri}/\mathcal{E}_{i,periH}$. For the MA interface,
$F_{\mathrm{MA}}$ includes both the top metal--air region and the vertical
sidewall region.

For the cross-section IDC model in a flip-chip environment, two voltage
excitation schemes can be considered, and the distinction between them is
subtle. In the unbiased opposite-plane model sketched in
Fig.~\ref{fig:scaling_factor_fc_models}(a), the capacitor metals define the
electrostatic bias, while the opposite-chip ground plane is included
geometrically but is not assigned a voltage excitation. This choice is
appropriate when the scaling factor is intended to describe the local edge
correction of the biased conductor pair without forcing the opposite ground
plane to share the same electrostatic reference. For example, it can be used to
describe the two metal pads of a floating transmon
\cite{seteFloatingTunableCoupler2021,yanayMediatedInteractionsNearest2022a}. In
the grounded opposite-plane model sketched in
Fig.~\ref{fig:scaling_factor_fc_models}(b), the opposite-chip ground plane is
tied to the same $0\,\mathrm{V}$ reference as one of the capacitor metal pads.
This choice is appropriate when one electrode of the IDC is galvanically
connected to the opposite ground plane. In that case, one electrode and the
opposite plane are treated together as ground, as is appropriate, for example,
for the metal island of an Xmon qubit
\cite{barendsCoherentJosephsonQubit2013b}.

In the first two columns of Fig.~\ref{fig:scaling_factor_fc_models}, we compare
the scaling factors obtained under these two voltage boundary conditions. The
curves in Fig.~\ref{fig:scaling_factor_fc_models}(c), (d), (f), and (g) show
that the two excitation schemes yield similar values of $F_{\mathrm{MS}}$ and
$F_{\mathrm{SA}}$. At $x_0=0.6\,\mu\mathrm{m}$, averaging over the simulated
$(w,g)$ values gives $F_{\mathrm{MS}}=7.9$ and $F_{\mathrm{SA}}=9.3$ for the
unbiased opposite-plane model, compared with $F_{\mathrm{MS}}=8.1$ and
$F_{\mathrm{SA}}=9.2$ for the grounded opposite-plane model. The strongest
boundary-condition dependence appears in $F_{\mathrm{MA}}$, shown in
Fig.~\ref{fig:scaling_factor_fc_models}(e) and (h). Grounding the opposite
plane reduces the average $F_{\mathrm{MA}}$ at $x_0=0.6\,\mu\mathrm{m}$ from
$90.0$ to $58.2$, corresponding to an average ratio of approximately $0.66$.
This trend is consistent with the physical picture that the opposite chip
primarily screens the field near the top metal surface and sidewall, whereas
the lower-chip metal--substrate and exposed substrate--air edge fields are much
less affected.

The third column of Fig.~\ref{fig:scaling_factor_fc_models} examines the
sensitivity of the results to the assumed dielectric constant of the lossy
surface layers. We use the same model as in
Fig.~\ref{fig:scaling_factor_fc_models}(a) but change the surface-layer
permittivity from $\epsilon_{\mathrm{layer}}=10$ to
$\epsilon_{\mathrm{layer}}=4$. At $x_0=0.6\,\mu\mathrm{m}$, this change yields,
on average, an $8\%$ lower $F_{\mathrm{MS}}$, a $34\%$ higher
$F_{\mathrm{SA}}$, and a $46\%$ lower $F_{\mathrm{MA}}$, as seen by comparing
panels (c), (d), and (e) with panels (i), (j), and (k). 
However, the prefactors in
Eqs.~\eqref{eq:p_ma}--\eqref{eq:p_sa} vary with the interface dielectric
constant in the opposite direction. 
Reducing $\epsilon_{\mathrm{layer}}$
therefore tends to increase the total participation ratios associated with the MA and
MS interfaces, while decreasing the total participation ratio associated with
the SA interface
\cite{quintanaCharacterizationReductionMicrofabricationinduced2014}. 
As a result, 
a lower $\epsilon_{\mathrm{layer}}$ generally leads to a larger
dielectric loss from the MA and MS interfaces. 
The actual dielectric constants of these interfaces depend on the materials and
fabrication processes used. 
In this work, we assume $\epsilon=10$ for all lossy
interfaces, following the common convention used in simulations of
aluminium-on-sapphire devices
\cite{wennerSurfaceLossSimulations2011a,wangSurfaceParticipationDielectric2015}.

The use of a constant scaling factor to convert $p_{i,periH}$ into $p_{i,peri}$
relies on the assumption that the electric field near a metal edge exhibits a
local scaling property that is insensitive to the distant electromagnetic
environment \cite{wangSurfaceParticipationDielectric2015}. Accordingly, the
simulated $F_i$ values for different sets of $w$ and $g$ should converge when
$x_0\ll w,g$. Figure~\ref{fig:scaling_factor_fc_models} shows that
$F_{\mathrm{MS}}$ and $F_{\mathrm{SA}}$ converge well as $x_0\rightarrow 0$,
whereas $F_{\mathrm{MA}}$ for different $(w,g)$ values does not exhibit a clear
convergence trend. Instead, $F_{\mathrm{MA}}$ shows a pronounced dependence on
the metal gap $g$, even when $x_0$ is much smaller than both $w$ and $g$. This
behavior may arise from the more complex electromagnetic environment near the
vertical metal sidewall \cite{wennerSurfaceLossSimulations2011a}.

For the final flip-chip participation analysis, we take
$x_0=0.6\,\mu\mathrm{m}$ and adopt rounded correction factors based primarily
on the unbiased opposite-plane model in
Fig.~\ref{fig:scaling_factor_fc_models}(a),
\begin{equation}
F_{\mathrm{MA}}=90,\qquad
F_{\mathrm{MS}}=8,\qquad
F_{\mathrm{SA}}=9 .
\end{equation}
In our devices, both unbiased opposite-plane and grounded opposite-plane
configurations are present. Because the grounded opposite-plane model yields a
smaller $F_{\mathrm{MA}}$, adopting $F_{\mathrm{MA}}=90$ may slightly
overestimate the participation of the MA interface. In addition, because the
calculated $F_{\mathrm{MA}}$ depends on the metal gap, a more accurate
treatment would further divide the MA perimeter region into large-gap and
small-gap regions and assign different scaling factors to them. Therefore, our
simulated $p_{\mathrm{ma}}$ may be less accurate than $p_{\mathrm{ms}}$ and
$p_{\mathrm{sa}}$. Nevertheless, the goal of this work is to compare the
surface participation ratios of different designs rather than to determine
their absolute values. For that purpose, the simulation method described above
is sufficient.

\section{Frequency-control check for opposite-chip hole designs}
\label{app_frequency_control}

The opposite-chip hole designs in Sec.~\ref{sec:parameter_dependence} 
shift the solved qubit-mode frequencies upward,
which could in principle change the mode hybridization and the extracted SPR.
To separate this frequency effect from the geometric effect of the hole, we
compare the baseline geometries with frequency-control variants
$\mathcal{O}_{50}^{\omega}$ and $\mathcal{O}_{80}^{\omega}$.  These controls
keep the same geometry as $\mathcal{O}_{50}$ and $\mathcal{O}_{80}$, respectively, 
while changing the lumped inductance $L_j$ in the eigenmode
simulation, shifting the qubit-mode frequency by about $17\%$.

\begin{figure}[h]
    \centering
    \includegraphics[width=0.9\columnwidth]{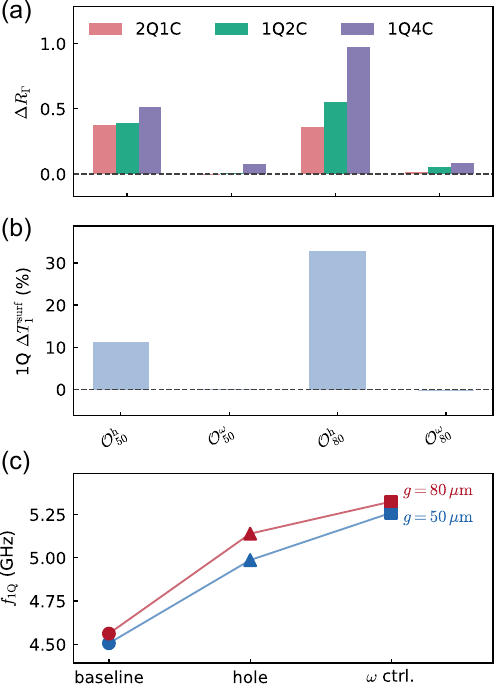}
    \caption{
    Frequency-control check for the opposite-chip-hole comparison.
    (a) Changes in $R_\Gamma$ for the qubit modes in 2Q1C, 1Q2C, and 1Q4C.  
    The hole bars compare
    $\mathcal{O}_{50}^{\rm h}$ with $\mathcal{O}_{50}$ and
    $\mathcal{O}_{80}^{\rm h}$ with $\mathcal{O}_{80}$; the frequency-control
    bars compare $\mathcal{O}_{50}^{\omega}$ with $\mathcal{O}_{50}$ and
    $\mathcal{O}_{80}^{\omega}$ with $\mathcal{O}_{80}$.  (b) Corresponding
    percentage changes in the isolated 1Q $T_1^{\rm surf}$.  (c) Solved
    isolated-qubit frequencies for the baseline, hole, and frequency-control
    designs.  
    The much smaller $R_\Gamma$ and
    $T_1^{\rm surf}$ responses in panels (a) and (b), despite this larger
    frequency shift, support the conclusion that the hole-induced SPR change
    is dominated by geometry.
    }
    \label{fig:frequency_control_appendix}
\end{figure}

The resulting changes in the connectivity-induced surface-loss penalty $R_\Gamma$ 
are much smaller than those produced by the hole geometries, as shown in
Fig.~\ref{fig:frequency_control_appendix}(a).  
For $\mathcal{O}_{50}^{\omega}$,
the 2Q1C and 1Q2C changes are below 0.01, while 1Q4C changes by 0.08.  
For $\mathcal{O}_{80}^{\omega}$, 
the corresponding changes are 0.01, 0.05, and 0.08.  
By contrast, adding the hole changes the 2Q1C, 1Q2C, and 1Q4C
penalties by 0.38, 0.39, and 0.51 for the $g=50\,\mu\mathrm{m}$ design, and
by 0.36, 0.55, and 0.97 for the $g=80\,\mu\mathrm{m}$ design.
The isolated-qubit lifetime shows the same separation
[Fig.~\ref{fig:frequency_control_appendix}(b)]:
the frequency controls change $T_1^{\rm surf}$ by only
$0.03\,\mu\mathrm{s}$ ($0.02\%$) and $-0.38\,\mu\mathrm{s}$ ($-0.27\%$),
whereas the hole geometries increase it by
$14.9\,\mu\mathrm{s}$ ($11.2\%$) and
$45.5\,\mu\mathrm{s}$ ($32.8\%$).
As shown in Fig.~\ref{fig:frequency_control_appendix}(c), the
frequency-control variants shift the isolated-qubit frequency by about
$16.7\%$, which is larger than the hole-induced shifts of about $10.7\%$ and
$12.7\%$ for the $g=50\,\mu\mathrm{m}$ and $g=80\,\mu\mathrm{m}$ designs,
respectively.  
The frequency controls in this appendix 
therefore support the conclusion that
the hole-induced changes in $T_1^{\rm surf}$ and $R_\Gamma$ are dominated by
geometry rather than by the shifted qubit-mode frequency, although frequency
and detuning can still affect the detailed composition of individual dressed
modes.

\section{Interface hierarchy and chip localization}
\label{app_simulation_detail}

\begin{table}[h]
    \centering
    \caption{
    Bulk-energy participation ratios of the qubit mode. 
    $p_{\rm subs,top}$ and $p_{\rm subs,bottom}$ denote the energy fractions in
    the top and bottom substrates, respectively.  The vacuum participation is
    obtained from
    $p_{\rm vac}=1-p_{\rm subs,top}-p_{\rm subs,bottom}$.
    $f_q$ refers to the qubit-mode frequency.
    }
    \label{tab:substrate_energy_partition}
    \begin{tabular}{llcccc}
        \hline\hline
        Design & model & $f_q$ (GHz) & $p_{\rm subs,top}$ &
        $p_{\rm subs,bottom}$ & $p_{\rm vac}$ \\
        \hline
        $\mathcal{O}_{50}$ & 1Q & 4.506 & 0.5833 & 0.0000 & 0.4167 \\
        $\mathcal{O}_{50}$ & 1Q4C & 4.220 & 0.6538 & 0.0003 & 0.3459 \\
        $\mathcal{O}_{50}^{\rm h}$ & 1Q & 4.987 & 0.7056 & 0.1244 & 0.1700 \\
        $\mathcal{O}_{50}^{\rm h}$ & 1Q4C & 4.553 & 0.7570 & 0.0892 & 0.1538 \\
        $\mathcal{O}_{80}$ & 1Q & 4.562 & 0.5602 & 0.0000 & 0.4398 \\
        $\mathcal{O}_{80}$ & 1Q4C & 4.242 & 0.6467 & 0.0003 & 0.3530 \\
        $\mathcal{O}_{80}^{\rm h}$ & 1Q & 5.139 & 0.6946 & 0.1411 & 0.1643 \\
        $\mathcal{O}_{80}^{\rm h}$ & 1Q4C & 4.604 & 0.7578 & 0.0925 & 0.1498 \\
        \hline\hline
    \end{tabular}
\end{table}

This appendix examines the interface-resolved and chip-resolved structure of
the participation ratios for the designs investigated in the main text.  
We decompose the simulated SPR into MA,
MS, and SA contributions and then separate each interface into top-chip and
bottom-chip components.  This view identifies which interfaces and which chip
surfaces carry most of the surface loss in the simulated geometries.

The detailed MA/MS/SA decompositions summarized in
Fig.~\ref{fig:simulation_detail_appendix} show that, 
for the displayed design families and model cases, 
both $p_{\rm MS}$ and $p_{\rm SA}$ are larger than $p_{\rm MA}$.  
This trend holds for the isolated 1Q, passive-claw 1QwP,
connected 1Q2C and 1Q4C cases, and also the coupler mode.
Thus the surface-loss estimate under the assumed loss tangents is controlled
primarily by MS and SA rather than by MA, consistent with 
previous surface-participation analyses in planar devices
~\cite{wennerSurfaceLossSimulations2011a,wangSurfaceParticipationDielectric2015}.

\begin{figure}[t]
    \centering
    \includegraphics[width=0.9\columnwidth]{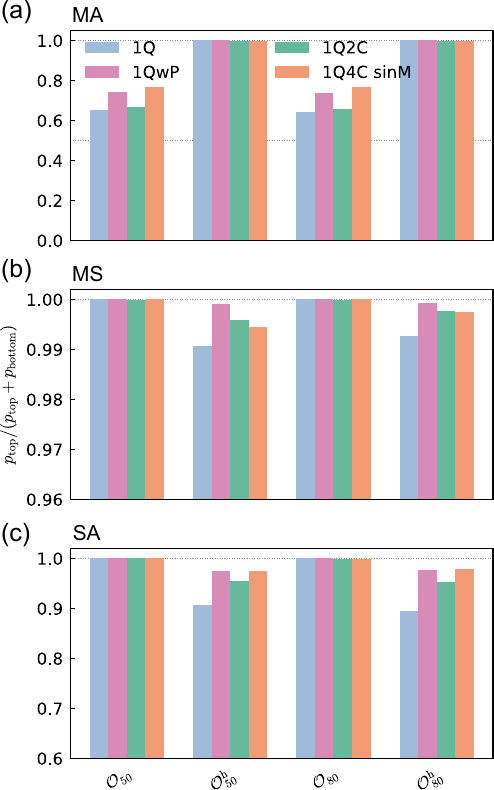}
    \caption{
    Top-chip fraction of the MA, MS, and SA participations for selected
    gap and opposite-chip-hole designs.  The plotted quantity is
    $p_{\rm top}/(p_{\rm top}+p_{\rm bottom})$ for each interface and model
    case.  Values near unity indicate that the corresponding participation is
    localized mainly on the top chip.
    }
    \label{fig:top_bottom_spr_appendix}
\end{figure}

To check which chip contributes most strongly to these interfaces, we
decompose the MA, MS, and SA participations into top-chip and bottom-chip
contributions for the gap and hole families.  As shown in
Fig.~\ref{fig:top_bottom_spr_appendix}, the MS and SA contributions are
strongly top-chip dominated.  The MA contribution is also larger on the top
chip for most checked cases, but its top-bottom asymmetry is weaker.  The
opposite-chip hole changes the absolute participation, 
but it does not overturn the top-chip-dominated trend 
for MS and SA interface.

As a complementary bulk-energy check, 
Table~\ref{tab:substrate_energy_partition}
summarizes the bulk electric-energy participation ratios 
in the top-chip substrate, bottom-chip substrate, 
and vacuum for the 1Q and 1Q4C models.  
The top-chip substrate stores the largest energy share in all
listed cases.  The unetched opposite ground plane provides an electrostatic
counterelectrode below the qubit and pulls part of the qubit-mode energy into
the vacuum gap between the two chips.  Etching a hole in this opposite plane
reduces that pull, 
so a larger fraction of the field energy remains inside the top substrate.  
This picture is consistent with the
increase of $p_{\rm subs,top}$ in the hole designs.  
Although this
bulk-energy participation is distinct from the surface participation used in the
loss model, it points to a related materials consideration: further reducing
dielectric loss may also require controlling bulk loss in the top substrate
~\cite{ganjamSurpassingMillisecondCoherence2024,
blandMillisecondLifetimesCoherence2025b}, 
in addition to improving the top-chip MS and SA interfaces.

\section{Capacitance and coupling parameters}
\label{app_qcq_capacitance}

In addition to the connectivity-induced penalty discussed in the main text,
the coupling strengths provide an important characterization of a
qubit-coupler lattice.  
This appendix therefore provides detailed capacitance parameters 
(see Table~\ref{tab:qcq_capacitance_appendix}) of
the designs investigated 
in Sec.~\ref{sec:connectivity_spr} and \ref{sec:parameter_dependence},
and uses them to calculate the corresponding coupling strengths and
anharmonicities following the method of
Refs.~\cite{liangTunablecouplingArchitecturesCapacitively2023,
marxerLongDistanceTransmonCoupler2023}.

The resulting parameters are summarized in
Table~\ref{tab:qcq_target_metrics_appendix}.  
The gap and hole controls mainly
act through the qubit capacitance.  
Increasing the qubit-ground gap from
$\mathcal{O}_{30}$ to $\mathcal{O}_{80}$ reduces $C_{Q,G}$ and increases the
average qubit-coupler coupling from about $63$ to $73\,\mathrm{MHz}$.  
Adding an opposite-chip hole reduces $C_{Q,G}$ more strongly, increases
$|\alpha_q|$, and raises $g_{qc}$ to about $86\,\mathrm{MHz}$ for
$\mathcal{O}_{50}^{\rm h}$ and $95\,\mathrm{MHz}$ for
$\mathcal{O}_{80}^{\rm h}$.  
By contrast, the claw controls primarily modify
the coupling network while leaving $\alpha_q$ nearly unchanged.  Reducing the
claw gap in $\mathcal{O}_{80}^{w-}$ weakens $g_{qc}$ and $g_{qq}$ relative to
$\mathcal{O}_{80}$, whereas shortening the claw in $\mathcal{O}_{80}^{d-}$
increases the extracted coupling strengths.
The IDC design $\mathcal{O}_{80}^{\rm IDC}$ provides the largest
claw-coupler capacitance and raises $g_{qc}$ to about $97\,\mathrm{MHz}$.
These results indicate that stronger coupling can, in some designs, come
at the expense of increased surface loss, as illustrated by
$\mathcal{O}_{80}^{\rm IDC}$.
However, this trade-off is not unavoidable: appropriate geometric
variations, such as $\mathcal{O}_{80}^{d-}$, can increase $g_{qc}$
while reducing the surface-loss penalty.

\begin{figure*}[p]
    \centering
    \includegraphics[
        width=0.8\textwidth,
        height=0.92\textheight,
        keepaspectratio
    ]{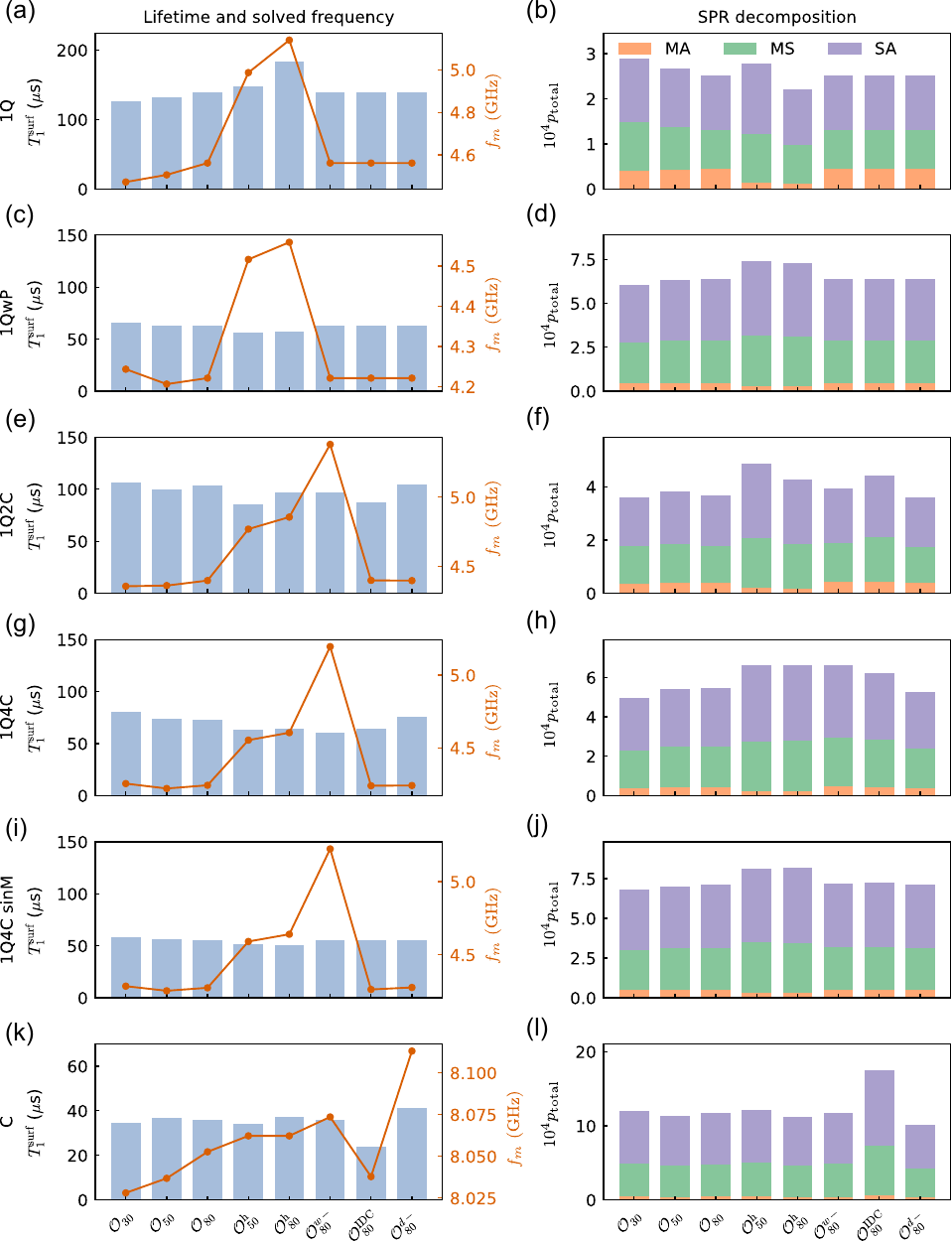}
    \caption{
    Detailed simulation diagnostics for the design investigated in the main text.  
    The left column in each model row shows the estimated $T_1^{\rm surf}$ together
    with the solved eigenmode frequency.  The right column shows the
    corresponding MA/MS/SA SPR decomposition, plotted as
    $10^4p_{\rm total}$.  The displayed model rows are 1Q, 1QwP, 1Q2C,
    1Q4C, 1Q4C sinM, and the coupler mode.
    }
    \label{fig:simulation_detail_appendix}
\end{figure*}

Table~\ref{tab:qcq_target_metrics_appendix} also lists the coupler off
frequencies $f_{c,\rm off}$, 
at which the qubit-coupler effective coupling is minimized.  
The solved coupler-mode frequencies in our SPR
simulations (see Fig.~\ref{fig:simulation_detail_appendix}(k)) are close to
these off frequencies.  Therefore, the surface-loss simulations in the main
text primarily probe operating points near the qubit-coupler off point, or
idle point, and should not be interpreted as a complete loss model for every
possible coupler bias during a gate.

\begin{table*}[h]
    \centering
    \caption{
    Capacitance parameters of the QCQ unit cell.  Values are capacitance
    matrix terms in fF for the qubit nodes $Q_1$ and $Q_2$, coupling-pad
    nodes $P_1$ and $P_2$, coupler node $C$, and ground node $G$.
    }
    \label{tab:qcq_capacitance_appendix}
    \begin{ruledtabular}
    \begin{tabular}{l c c c c c c c c c}
        design & $C_{Q1,G}$ & $C_{Q2,G}$ & $C_{C,G}$ & $C_{Q1,P1}$ & $C_{Q2,P2}$ & $C_{P1,C}$ & $C_{P2,C}$ & $C_{P1,P2}$ & $C_{Q1,Q2}$ \\
        \hline
        $\mathcal{O}_{30}$ & 77.54 & 77.49 & 49.53 & 6.40 & 6.36 & 39.44 & 41.02 & 4.07 & 0.00 \\
        $\mathcal{O}_{50}$ & 74.87 & 74.90 & 49.56 & 6.63 & 6.76 & 39.19 & 40.68 & 4.04 & 0.00 \\
        $\mathcal{O}_{50}^{\rm h}$ & 59.49 & 59.55 & 49.46 & 7.67 & 7.67 & 39.38 & 40.95 & 4.08 & 0.01 \\
        $\mathcal{O}_{80}$ & 73.53 & 73.59 & 49.49 & 7.04 & 7.09 & 39.07 & 40.71 & 4.04 & 0.01 \\
        $\mathcal{O}_{80}^{\rm h}$ & 55.86 & 55.83 & 49.23 & 8.27 & 8.18 & 38.03 & 39.53 & 4.01 & 0.01 \\
        $\mathcal{O}_{80}^{w-}$ & 73.60 & 73.61 & 49.68 & 6.96 & 6.99 & 38.96 & 40.58 & 3.99 & 0.01 \\
        $\mathcal{O}_{80}^{\rm IDC}$ & 73.32 & 73.29 & 39.65 & 7.34 & 7.35 & 71.79 & 73.09 & 4.92 & 0.01 \\
        $\mathcal{O}_{80}^{d-}$ & 73.36 & 73.50 & 49.44 & 7.24 & 7.24 & 39.07 & 40.56 & 4.01 & 0.01 \\
    \end{tabular}
    \end{ruledtabular}
\end{table*}

\begin{table*}[h]
    \centering
    \caption{
    Qubit-coupler parameters obtained from the capacitance matrix after tuning the
    qubit and coupler frequencies to $f_1=f_2=4.5\,\mathrm{GHz}$ and
    $f_c=8.0\,\mathrm{GHz}$, respectively.
    The off point $f_{c,\rm off}$ is the coupler
    frequency that minimizes the qubit-coupling effective XY coupling 
    at fixed $f_q=4.5\,\mathrm{GHz}$.
    The averaged qubit-coupler coupling is defined as $g_{qc}=(g_{1c}+g_{2c})/2$.
    Couplings and anharmonicities are in MHz;
    Josephson energies and $f_{c,\rm off}$ are in GHz.
    }
    \label{tab:qcq_target_metrics_appendix}
    \begin{ruledtabular}
    \begin{tabular}{l c c c c c c c c c}
        design & $g_{1c}$ & $g_{2c}$ & $g_{qc}$ & $g_{qq}$
        & $\alpha_q$ & $\alpha_c$ & $E_{Jq}$ & $E_{Jc}$
        & $f_{c,\rm off}$ \\
        \hline
        $\mathcal{O}_{30}$ & 63.5 & 63.1 & 63.3 & 1.40 & -240.8 & -182.7 & 13.09 & 48.16 & 8.437 \\
        $\mathcal{O}_{50}$ & 67.8 & 68.7 & 68.3 & 1.63 & -248.1 & -183.9 & 12.78 & 47.90 & 8.391 \\
        $\mathcal{O}_{50}^{\rm h}$ & 86.4 & 86.1 & 86.3 & 2.61 & -306.5 & -183.6 & 10.91 & 47.95 & 8.415 \\
        $\mathcal{O}_{80}$ & 72.8 & 73.1 & 72.9 & 1.88 & -251.4 & -184.3 & 12.65 & 47.80 & 8.313 \\
        $\mathcal{O}_{80}^{\rm h}$ & 95.3 & 94.3 & 94.8 & 3.22 & -323.7 & -186.9 & 10.49 & 47.20 & 8.164 \\
        $\mathcal{O}_{80}^{w-}$ & 66.4 & 66.8 & 66.6 & 1.60 & -251.4 & -181.1 & 12.65 & 48.56 & 8.150 \\
        $\mathcal{O}_{80}^{\rm IDC}$ & 97.2 & 96.2 & 96.7 & 2.81 & -251.2 & -160.0 & 12.66 & 54.37 & 10.988 \\
        $\mathcal{O}_{80}^{d-}$ & 82.0 & 81.4 & 81.7 & 2.52 & -251.2 & -187.4 & 12.66 & 47.08 & 7.733 \\
    \end{tabular}
    \end{ruledtabular}
\end{table*}



\clearpage


\bibliography{reference}

\end{document}